\documentclass[letterpaper,preprint,showpacs,preprintnumbers,amsmath,ammsymb]{revtex4}

\usepackage{color}
\usepackage{graphicx}
\usepackage{dcolumn}
\usepackage{bm}
\usepackage{amssymb,amsfonts,amsmath}
\usepackage{graphicx}
\usepackage{threeparttable}

\begin{document}


\title{\bf{Post density functional theoretical
    studies of highly polar semiconductive
    Pb(Ti$_{1-x}$Ni$_{x}$)O$_{3-x}$ solid solutions: The effects of
    cation arrangement on band gap } \\[11pt] }

\author{G. Y. Gou$^{1}$, J. W. Bennett$^{2}$, H. Takenaka$^{1}$ and A. M. Rappe$^{1}$}
\affiliation{
  1. The Makineni Theoretical Laboratories, Department of
  Chemistry, University of Pennsylvania, Philadelphia, PA
  19104-6323, USA\\
  2. Department of Physics and Astronomy, Rutgers University, Piscataway, New Jersey 08854-8019, USA\\}

\date{\today}

\begin{abstract}

We use a combination of conventional density functional theory (DFT)
and post-DFT methods, including the local density approximation plus
Hubbard $U$ (LDA+$U$), PBE0, and self-consistent $GW$ to study the
electronic properties of Ni-substituted PbTiO$_{3}$ (Ni-PTO) solid
solutions. We find that LDA calculations yield unreasonable band
structures, especially for Ni-PTO solid solutions that contain an
uninterrupted NiO$_{2}$ layer. Accurate treatment of localized
states in transition-metal oxides like Ni-PTO requires post-DFT
methods. $B$-site Ni/Ti cation ordering is also investigated. The
$B$-site cation arrangement alters the bonding between Ni and O, and
therefore strongly affects the band gap ($E_{\rm g}$) of Ni-PTO. We
predict that Ni-PTO solid solutions should have a direct band gap in
the visible light energy range, with polarization similar to the
parent PbTiO$_{3}$. This combination of properties make Ni-PTO solid
solutions promising candidate materials for solar energy conversion
devices.

\end {abstract}


\maketitle

\section*{\label{sec:level1} Introduction}

The bulk photovoltaic effect (BPVE) has been observed in
ferroelectric perovskite oxides such as
Pb(Zr$_{1-x}$Ti$_{x}$)O$_{3}$~\cite{Yarmarkin00p511,Uprety07p084107,Inoue86p2809},
LiNbO$_{3}$ ~\cite{Glass74p233} and BaTiO$_{3}$ ~\cite{Brody73p673}.
This effect has attracted increasing research interest, in view of
its potential application for solar energy conversion devices. As
incident light is absorbed by a single-phase ferroelectric material,
the photo-excited electrons and holes are spontaneously separated,
preventing recombination, and producing a photovoltaic effect. One
ferroelectric material extensively studied for this purpose is
BiFeO$_{3}$. It has a strong polarization $\approx$ 0.9 C/m$^{2}$
~\cite{Wang03p1719,Neaton05p014113} and a direct band gap $\approx$
2.67 eV ~\cite{Basu08p091905}. Recent experiments have demonstrated
the BPVE of BiFeO$_{3}$ under visible incident light
~\cite{Hauser08p222901, Yang09p062909, Choi09p63}. However,
BiFeO$_{3}$ is one of very few ferroelectric materials that have
both large spontaneous polarization and small $E_{\rm g}$. Most
solid oxide ferroelectrics are wide-gap insulators ($E_{\rm g}$$>$
3.0 eV) that absorb very little of the visible spectrum.

Recent theoretical work on Pb($M$$_{x}$Ti$_{1-x}$)O$_{3-x}$
perovskite solid solutions ($M$=Ni, Pd and Pt) provides a promising
strategy for designing highly polar semiconducting
oxides~\cite{Bennett08p17409,Bennett10p184106}. By substituting the
$B$-site Ti with an O-vacancy-stabilized $d^{8}$ $M^{2+}$ cation,
the resulting system displays a decreased $E_{\rm g}$ when
compared to PbTiO$_{3}$ (PTO), while retaining a large polarization.
Of these proposed materials, Ni-PTO solid solutions possess great
potential for practical solar applications, as they can be
synthesized from relatively inexpensive raw materials. However, due
to the strongly correlated Ni-3$d$ electrons, Ni-PTO has a complex
electronic structure that is not captured by standard LDA
calculations~\cite{Anisimov91p943}. In this study, we explore the
electronic structure of Ni-PTO solid solutions with post-DFT
calculations, including LDA+$U$, PBE0 hybrid functional and
self-consistent $GW$. We predict the band gap and other electronic
properties for Ni-PTO solid solutions of various compositions and
$B$-site orderings, to guide future experimental synthesis and
measurement.

\section*{\label{sec:level1} Methodology}

We perform first-principles calculations with a plane-wave basis
set, as implemented in  the Quantum-Espresso
~\cite{Giannozzi09p395502} and ABINIT~\cite{Gonze02p478} codes. The
LDA exchange-correlation functional is used for structural
relaxations, with a $6\times6\times6$ Monkhorst-Pack k-point grid
~\cite{Monkhorst76p5188} and a 50 Ry plane-wave cutoff. All atoms
are represented by norm-conserving ~\cite{Rappe90p1227} optimized
nonlocal~\cite{Ramer99p12471} pseudopotentials, generated with the
OPIUM code~\cite{Opium}. The electronic contribution to the
polarization is calculated following the Berry's phase formalism.
~\cite{Resta94p899,Kingsmith93p1651}

Several supercell configurations are used, including
1$\times$1$\times$2 and 1$\times$1$\times$3 layered structures (with
$B$-cation alternation along (001) direction); rocksalt
(Pb$_{2}$NiTiO$_{5}$), $\sqrt{2}\times\sqrt{2}\times$2 and
2$\times$2$\times$2 structures (Ni-O$_{5}$ cages are separated by
Ti-O$_{5}$ cages). By studying these configurations, the effects of
different compositions and $B$-site cation orderings can be
investigated. Post-DFT methods are then applied to these Ni-PTO
systems. In the next section, we briefly describe each post-DFT
method, to facilitate comparison of results obtained with each
method.

\section*{\label{sec:level1} Results and Discussion}
\subsection*{\label{sec:level2} Post-DFT methods}

Before applying post-DFT methods to Ni-PTO, we study its parent
materials, tetragonal PbTiO$_{3}$ and cubic NiO, as reference
systems to evaluate the performance of post-DFT methods for related
materials. We do so because it is known that the Kohn-Sham (KS) band
gap determined by the conventional LDA method is systematically
underestimated by 50-100 \%
~\cite{Hybertsen85p1418,Hybertsen86p5390}. To improve the quality of
the electronic structure description of Ni-PTO, especially the
treatment of correlation for the localized Ni-3$d$ orbitals, we
employ LDA+$U$, the PBE0 hybrid functional, and self-consistent $GW$
(sc-$GW$) methods.

\subsubsection* {\label{sec:level3} LDA+$U$}

The LDA+$U$ method is a computationally inexpensive way to improve
the DFT-LDA band gap. A simplified version of the rotationally
invariant formulation of the LDA+$U$ method~\cite{Anisimov91p943} is
employed in the present work, where $U$ is assigned to be the value
of the spurious curvature of the total energy of the system with
respect to the variation in the number of electrons ($n_{i}$) in a
localized orbital. ~\cite{Cococcioni05p035105} The effective
interaction parameter $U$ is determined by the change of orbital
occupations $n_{i}$ with respect to an external potential
$\alpha_{j}$ projected onto the corresponding localized level:

\begin{equation}
U=(\chi_{0}^{-1}-\chi^{-1})_{ii},\quad
\chi_{ij}=\frac{dn_{i}}{d\alpha_{j}},
\end{equation}\

Here the $\chi_{ij}$ are the elements of the linear response matrix,
$n_{i}$ is the occupation number of the localized levels at site $i$
and $\alpha_{j}$ represents the potential shift applied on the
localized orbital at site $j$. $\chi$ (the screened response matrix)
includes all screening effects from the crystal environment which is
associated with the localized electrons, and $\chi_{0}$ (the
unscreened response matrix) contains non-screening effects of the
total energy of the non-interacting Kohn-Sham states associated with
the system. The latter term can be computed from the first iteration
in the self-consistent cycle.

\begin{figure}[t]
\centering
\includegraphics[width=8cm]{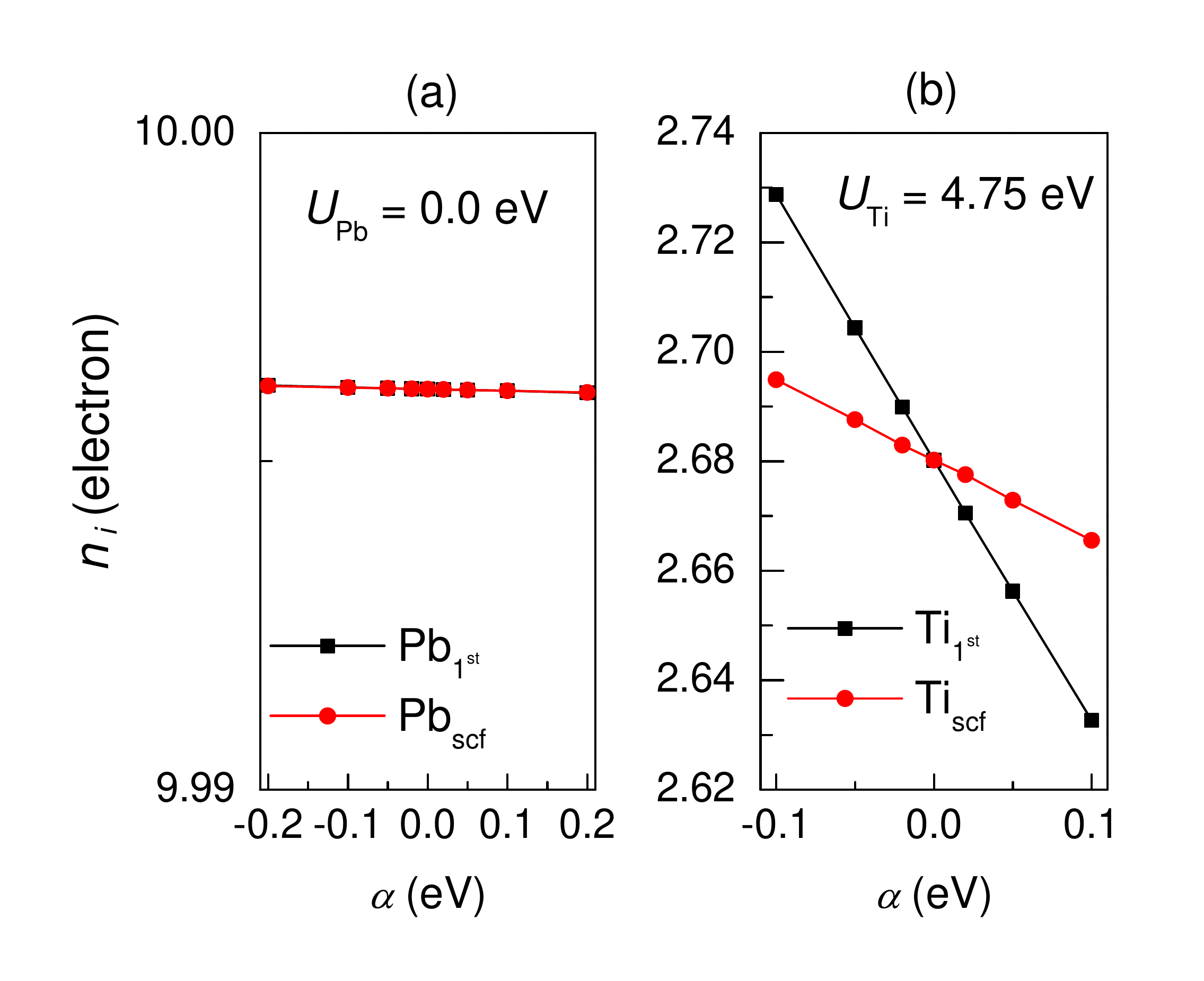}
\caption{(Color online) The occupation number ($n_{i}$) of localized
$d$-orbitals for Pb and Ti in PbTiO$_{3}$, with respect to potential
shift $\alpha$. Shown as (black) squares are the values obtained
using self-consistent (scf) relaxation, and shown as (red) circles
are those without (1$\rm ^{st}$), for constructing $\chi$ and
$\chi_{0}$ respectively.} \label{Fig.1}
\end{figure}

Tetragonal PbTiO$_{3}$ is selected as an example for detailed study.
A 2$\times$2$\times$2 PbTiO$_{3}$ supercell is used, because it is
large enough to give a converged linear response calculation of $U$.
Fig.\ 1 displays the calculated $U$ values for Pb and Ti atoms of
PbTiO$_{3}$. In the rotationally invariant LDA+$U$ method, on-site
interaction energy
$E_{U}$$\propto$$U$$\cdot$[$\lambda$(1-$\lambda$)]. For fully
occupied orbitals such as Pb 5$d$, $\lambda$= 1, so on-site
correction plays no effect. As displayed in Fig.\ 1 (a), the number
of electrons on the Pb 5$d$ orbitals will not change with respect to
the external potential $\alpha$, so $U$=0 eV. Ti has partially
filled 3$d$ orbitals, so it has a non-zero $U$ value.

\begin{table}
\setlength{\tabcolsep}{0.15cm}
\renewcommand{\arraystretch}{1.2}
\caption{Hubbard $U$ terms (in eV) for Ni and Ti in PbTiO$_{3}$, NiO
and  Ni-PTO solid solutions, obtained with the linear-response
approach.}
\begin{tabular}{c|cc}
\hline\hline
Composition &  $U_{\rm{Ti}}$ &  $U_{\rm{Ni}}$   \\
\hline
PbTiO$_{3}$                                         & 4.75 &  -   \\
NiO                                                 &   -  & 5.70 \\
layered Pb$_{2}$NiTiO$_{5}$                         & 4.85 & 9.31 \\
rocksalt Pb$_{2}$NiTiO$_{5}$                        & 4.85 & 8.92 \\
1$\times$1$\times$3 Pb$_{3}$NiTi$_{2}$O$_{8}$                     & 4.75 & 9.35 \\
$\sqrt{2}\times\sqrt{2}\times$2 Pb$_{4}$NiTi$_{3}$O$_{11}$  & 4.75 & 8.86 \\
2$\times$2$\times$2 Pb$_{8}$NiTi$_{7}$O$_{23}$                    & 4.77 & 8.91 \\
\hline\hline
\end{tabular}
\end{table}

Following the procedure above, using the self-consistent linear
response method, as implemented in Quantum-Espresso, the Hubbard $U$
terms for Ti and Ni in NiO, PTO, and all Ni-PTO solid solutions can
be determined (see in Table I).

\subsubsection* {\label{sec:level1} PBE0 hybrid density functional
with exact-exchange}

\begin{figure}[t]
\centering
\includegraphics[width=8.5cm]{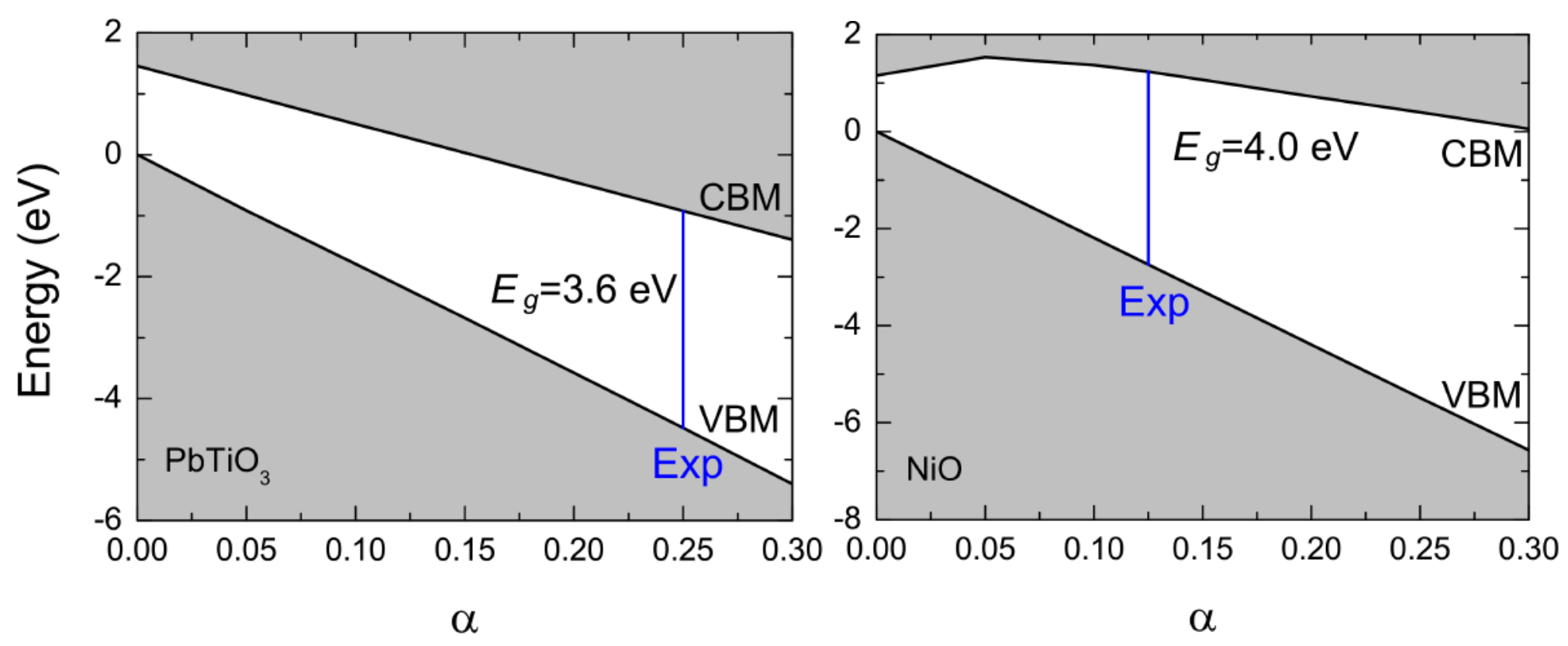}
\caption{(Color online) Valence band maximum (VBM) and conduction
band minimum (CBM) versus PBE0 functional exact-exchange fraction
$\alpha$ for PbTiO$_{3}$ and NiO. Vertical lines represent the
experimental band gaps, illustrating how $\alpha_{0}$ is chosen (see
Table II).} \label{Fig.2}
\end{figure}

DFT-LDA does not include exact exchange. This is unlike the PBE0
functional, which includes a linear combination of exact-exchange and
GGA-PBE exchange ~\cite{Perdew96p9982}. The electron correlation term
is the same as that of PBE functional ~\cite{Perdew96p3865}:

\begin{equation}
E_{xc}^{\rm{PBE0}}=\alpha
E_{x}+(1-\alpha)E_{x}^{\rm{PBE}}+E_{c}^{\rm{PBE}}.
\end{equation}

\noindent where $\alpha$ is the fraction of exact exchange. As PBE0
includes a portion exact exchange, it can greatly reduce the
self-interaction error of LSDA or GGA functional, and therefore
greatly improves the calculated band gap for most semiconductors and
insulators. ~\cite{Perdew96p3865} However, due to the long-range
nature of the exact-exchange interaction and resultant large
computational cost, (about one to two orders of magnitude more than
LDA in the present work) PBE0 calculations are mainly applied to
periodic systems with small unit cells, such as those described in
the present study. ~\cite{Paier06p154709}

In the original PBE0 functional, $\alpha$=0.25. However, it is
argued that there is no universal $\alpha$ that is applicable for
all materials ~\cite{Alkauskas08p106802}. In order to predict band
gap values for new materials more accurately, it is considered
useful to select an $\alpha$ that yields accurate band gaps for the
relevant prototype systems. Fig.\ 2 shows the $\alpha$ dependence of
the conduction and valence band edges of PbTiO$_{3}$ and NiO.
Comparing with experimental band gap values, we obtain an optimal
$\alpha_{0}$=0.25 for PbTiO$_{3}$ and 0.125 for NiO (Fig.\ 2 and
Table II).

\begin{table}
\setlength{\tabcolsep}{0.15cm}
\renewcommand{\arraystretch}{1.2}
\caption{Band gaps (in eV) of PbTiO$_{3}$ and NiO, calculated using
LDA, PBE0 (with the optimal fraction of exact exchange $\alpha_{0}$)
and sc-$GW$ methods.}
\begin{tabular}{ccccc}
\hline\hline
            &  LDA  &   PBE0 ($\alpha_{0}$) &  sc-$GW$    & Exp.    \\
\hline
PbTiO$_{3}$ &  1.46 &   3.50 (0.25)         &  3.82       & 3.60 ~\cite{Shirane56p131}  \\
NiO         &  1.16 &   3.98 (0.125)        &  4.14       & 4.0-4.3 ~\cite{Cheetham83p6964, Sawatzky84p2339}  \\
\hline\hline
\end{tabular}
\end{table}

\subsubsection* {\label{sec:level1} Self-consistent $GW$}

It is widely believed that the self-consistent $GW$ method is the
most accurate way to obtain electronic band structures, since it
includes both the non-locality and dynamical correlations absent in
DFT-LDA. Therefore, obtaining the self energy, $\Sigma$, is the key
quantity of any $GW$ calculation. It was shown by Hedin {\em et al.}
that the self-energy can be represented as a product of the Green's
function $G$ and the screened Coulomb interaction $W$,
$\Sigma=iGW$.~\cite{Hedin65p796}

Since our system has a complicated electronic band structure, in
which the band gap is dictated by the bonding interactions between
the correlated Ni 3$d$-states and O 2$p$-states, the self-consistent
$GW$ (sc-$GW$) method probably gives the most accurate
quasi-particle (QP) electronic structures of Ni-PTO.

Following the methodology developed by Lebegue {\em et al.},
~\cite{Lebegue03p155208} we employ the contour deformation
method~\cite{Bruneval06p045102, Anisimovbook}, as implemented in
ABINIT, to perform the numerical integral for the self-energy.
Within this scheme, the contour of the frequency integral for
self-energy is deformed into integrals along the imaginary and real
axes. The numerical integral along the imaginary axis can be
evaluated by Gaussian quadrature. We find that integrating up to 114
eV and using 10 quadrature points yields converged results for the
imaginary axis. The real axis integral is calculated by summing
values of the Coulomb screening at a uniform mesh of frequencies,
with a mesh spacing of 1.25 eV, from 0 to 120 eV.

This upper limit of real axis integration is chosen from spectral
function data, ~\cite{Bruneval06p045102,Anisimovbook} (electron
energy loss data in the present work), since the correlation part of
the self-energy can be represented by a spectral function.

Due to the expensive computational cost (three to four orders of
magnitude more than DFT-LDA in the present work), we restrict
sc-$GW$ calculations to only one Ni-PTO system -- rocksalt
structure. For this configuration, 400 bands are included, with a
plane-wave cutoff of 25 Ha for the QP and LDA states, and a
$2\times2\times2$ Monkhorst-Pack k-point
grid.~\cite{Monkhorst76p5188} As LDA calculations can qualitatively
predict a correct semiconducting ground state for rocksalt Ni-PTO
(Fig.\ 6), we choose to represent the initial QP wave functions in
the basis set of LDA wave functions. QP band gaps are converged to
within 0.05 eV.

\subsection*{\label{sec:level2} Comparison of results for parent end members}

\begin{figure*}
\centering
\includegraphics[width=12cm]{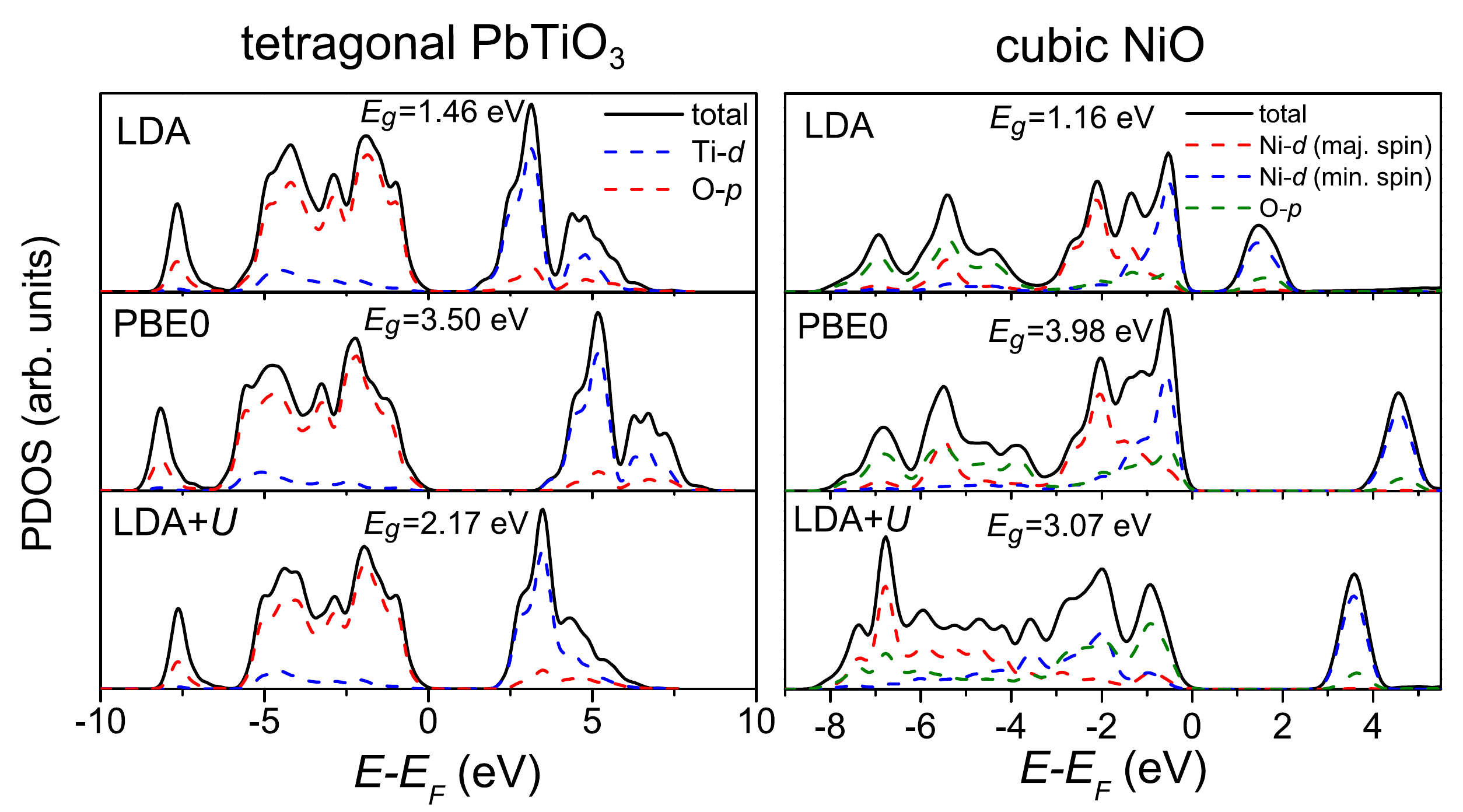}
\caption{(Color online) Band gap values and orbital-resolved PDOS of
tetragonal PbTiO$_{3}$ and cubic NiO using LDA, PBE0 and LDA+$U$. For
PBE0 calculations, the optimal $\alpha_{0}$ values are used.}
\label{Fig.3}
\end{figure*}

Fig.\ 3 shows the projected density of states (PDOS) for tetragonal
PbTiO$_{3}$ and cubic NiO, calculated with the LDA, PBE0 and LDA+$U$
methods, from which we can evaluate the accuracy of each post-DFT
method. Experimental studies show that tetragonal PbTiO$_{3}$ is an
insulator with an indirect band gap (X$\rightarrow$$\Gamma$) of 3.60
eV.  ~\cite{Shirane56p131} LDA severely underestimates this band gap
value by more than 50\%. The LDA+$U$ method gives a better band gap
value than LDA, but is still lower than PBE0, which comes closest to
experiment. This is because the exact-exchange in PBE0 is applied to
all states, whereas the Hubbard $U$ of LDA+$U$ acts directly {\em
only} on the localized Ti-3$d$ orbitals. Also, when compared to
PBE0, LDA+$U$ significantly alters the PDOS for PbTiO$_{3}$,
especially for Ti-3$d$ and O-2$p$ states.

NiO is an anti-ferromagnetic insulator with an experimentally measured
band gap between 4.0 and 4.3 eV ~\cite{Cheetham83p6964,
Sawatzky84p2339}. As shown in Fig.\ 3, LDA predicts a band gap of 1.16
eV and LDA+$U$ predicts 3.07~eV, while PBE0 (with $\alpha_{0}$=0.125)
reproduces the experimental band gap. For both LDA and PBE0
functionals, the PDOS has a clear split between majority spin Ni
t$_{2g}$ and e$_{g}$ orbitals, while the band gap of the system falls
between occupied minority-spin t$_{2g}$ and empty e$_{g}$
orbitals. Applying the Hubbard $U$ to the Ni-3$d$ orbitals increases
the NiO band gap, relative to LDA. However, LDA+$U$ predicts Ni-$d$
spectral weight located at a relatively low energy (2 -- 8 eV below
$E_{F}$). This is in contrast to both experimental photoemission
spectra ~\cite{Schuler05p115113} and theoretical dynamical mean field
theory (DMFT) results ~\cite{Kunes07p156404}. These studies suggest
that there should be a strong Ni-$d$ contribution to the valence band
of the system. Though LDA+$U$ does predict a larger band gap than
DFT-LDA, it does not reproduce the salient aspects of the experimental
valence-state spectrum of NiO.

\subsection*{\label{sec:level2} Electronic properties of Pb$_{2}$NiTiO$_{5}$ solid solutions}

We begin our study of Ni-PTO with the smallest supercell
configuration: Pb$_{2}$NiTiO$_{5}$. In this solid solution, half of
the Ti$^{4+}$ cations are replaced by O-vacancy-stabilized Ni$^{2+}$
cations~\cite{Bennett08p17409}.  This polar substitution will
generate a $V_{\rm O}^{\cdot\cdot}$ and $\rm Ni_{\rm
Ti}^{\prime\prime}$ defect pair (in Kroger-Vink notation
~\cite{Kroger56p273}), where the lowest-energy configuration
corresponds to the local dipole parallel to the overall polarization
($\vec{P}$). The Ni-PTO structure is tetragonal, with a large
polarization similar to the parent PbTiO$_{3}$ (Table IV). After
replacing Ti$^{4+}$ with Ni$^{2+}$, the remaining apical O will move
away from Ni$^{2+}$, making Ni coordination nearly square planar.
Since this O atom shifts opposite to the cations, it further
increases $\vec{P}$. For Pb$_{2}$NiTiO$_{5}$, two different $B$-site
cation orderings are studied: (i) 1$\times$1$\times$2 layered and
(ii) rocksalt Ni-PTO.

Layered Ni-PTO is structurally similar to the planar nickelate
SrNiO$_{2}$. A previous theoretical study of SrNiO$_{2}$ predicted it
to be a diamagnetic insulator with fully occupied $d_{xz}$, $d_{yz}$,
$d_{xy}$ and $d_{z^{2}}$ orbitals, leaving $d_{x^{2}-y^{2}}$
empty. ~\cite{Anisimov99p7901} Accordingly, since layered Ni-PTO is a
nickelate with planes of Ni$^{2+}$ ions, we anticipate that it should
have an insulating ground state.

Fig.\ 4 (a) shows the band structure for layered Ni-PTO along
high-symmetry directions in the Brillouin zone (BZ). The LDA band
structure reveals a two-dimensional character: the energy bands are
highly dispersive in the $k_{x}$-$k_{y}$ plane, but show a very weak
$z$-axis dispersion along $\Gamma$-$Z$. One strongly dispersive band
(dashed line in Fig.\ 4 (a)) crosses $E_{F}$, making the system
metallic in LDA. This band is chiefly composed of
Ni-3$d_{x^{2}-y^{2}}$ and O-2$p$. Isosurfaces for this band at
$\Gamma$, X and M points are displayed in Fig.\ 4 (b). There is a
pronounced hybridization between Ni-$d_{x^{2}-y^{2}}$ and O-$p$, as
the lobes from Ni-$d_{x^{2}-y^{2}}$ are directed towards the four O
ions. Specifically, the wave functions at X and M consist of
O($p_{x}$)-Ni($d_{x^{2}-y^{2}}$)-O($s$+$p_{z}$) and
O($p_{x}$)-Ni($d_{x^{2}-y^{2}}$)-O($p_{y}$) orbital combinations
respectively, both of which display anti-bonding character. However,
there is a strong bonding profile between Ni-$d_{x^{2}-y^{2}}$ and
O-$p_{z}$ at the $\Gamma$ point (Ni red (blue) to O red (blue)
lobes), which indicates that electron density is shared between
O-2$p$ and Ni-$d_{x^{2}-y^{2}}$ orbitals, again according to LDA
calculations.

\begin{figure}[t]
\centering
\includegraphics[width=8cm]{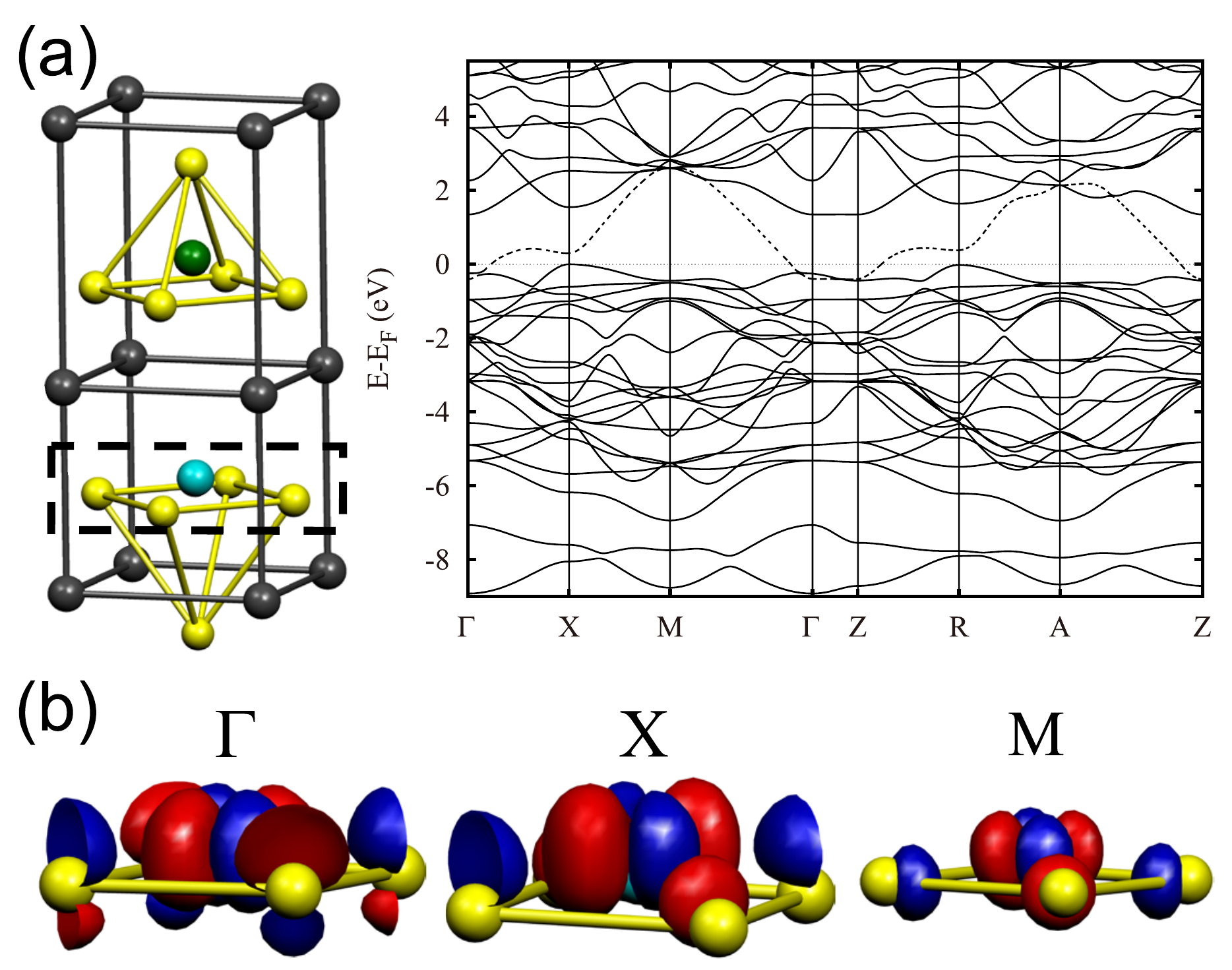}
\caption{(Color online) (a) Atomic structure of layered
Pb$_{2}$NiTiO$_{5}$, and its DFT-LDA calculated electronic band
structure. The band that crosses the Fermi level is shown as a
dashed line. (b) The wave function isosurfaces are plotted for this
band at $\Gamma$, $\rm X$ and $\rm M$. For clarity, only the
isosurface near the Ni-O$_{2}$ plane of the band indicated by the
dashed lines in (a) is plotted. The isosurface values are $\pm$ 0.04
e/\AA$^{3}$ (red and blue surfaces). The plots illustrate which
atomic orbitals compose each eigenstate: $\Gamma$:
O($p_{z}$)-Ni($d_{x^{2}-y^{2}}$)-O($p_{z}$), X:
O($p_{x}$)-Ni($d_{x^{2}-y^{2}}$)-O($s$+$p_{z}$) and M:
O($p_{x}$)-Ni($d_{x^{2}-y^{2}}$)-O($p_{y}$). This series of images
shows the drastic change in bonding from $\Gamma$ to X to M. Pb, Ti,
Ni and O atoms are represented as gray, green, cyan and yellow
spheres respectively.} \label{Fig.4}
\end{figure}

LDA calculations show that for layered Ni-PTO, the band gap is zero.
We find that in LDA calculations, the energy level of the
Ni-$d_{x^{2}-y^{2}}$ orbital is quite low, permitting this low-lying
orbital to backbond with O-$p_{z}$ orbitals. This enhancement of the
bonding interaction between Ni-3$d_{x^{2}-y^{2}}$ and O-2$p$ causes
LDA to predict a metallic ground state for layered Ni-PTO.

In our LDA+$U$ calculations, applying the Hubbard $U$ term (determined
to the Ni-3$d$ orbitals, in a self-consistent linear response
calculation~\cite{Cococcioni05p035105}), yields a split between
occupied and unoccupied Ni-3$d$ states. This provides a better
description of $E_{\rm g}$ for Ni-PTO, relative to LDA. However, the
PBE0 hybrid density functional, which includes exact-exchange energy,
can significantly improve the description of $E_{\rm g}$ of
ferroelectric oxides ~\cite{Blic08p165107}. To obtain the most
realistic results from a PBE0 calculation, the optimal exact-exchange
mixing fraction $\alpha$ from relevant prototype systems should be
used. Unlike PbTiO$_{3}$, with $E_{\rm g}$ between occupied O-2$p$ and
empty Ti-3$d$ states, both Ni-PTO and NiO have $E_{\rm g}$ between
occupied and unoccupied Ni-$d$ states.  Therefore, we use NiO as our
prototype parent system of Ni-PTO, and select $\alpha$=0.125 for PBE0
calculations of Ni-PTO. A comparison of the eigenvalues at $\Gamma$
obtained for the Ni 3$d$-states using LDA, LDA+$U$ and PBE0 is
presented in the supplementary material (EPAPS number).

Though the most computationally demanding of our post-DFT methods,
the many-body $GW$ approach is considered the most reliable first
principles method to obtain electronic band structure, and it has
been successfully applied to many kinds of oxides.
~\cite{Bruneval06p045102}. In our present work, we also employ this
method for the ground state calculation of rocksalt Ni-PTO,
obtaining a band gap of 1.83~eV. PBE0 predicts a value of 1.71~eV,
improving upon the results obtained by LDA (0.69~eV) and LDA+$U$
(1.59~eV), much closer to the sc-$GW$ result.

\begin{figure}[t]
\centering
\includegraphics[width=9cm]{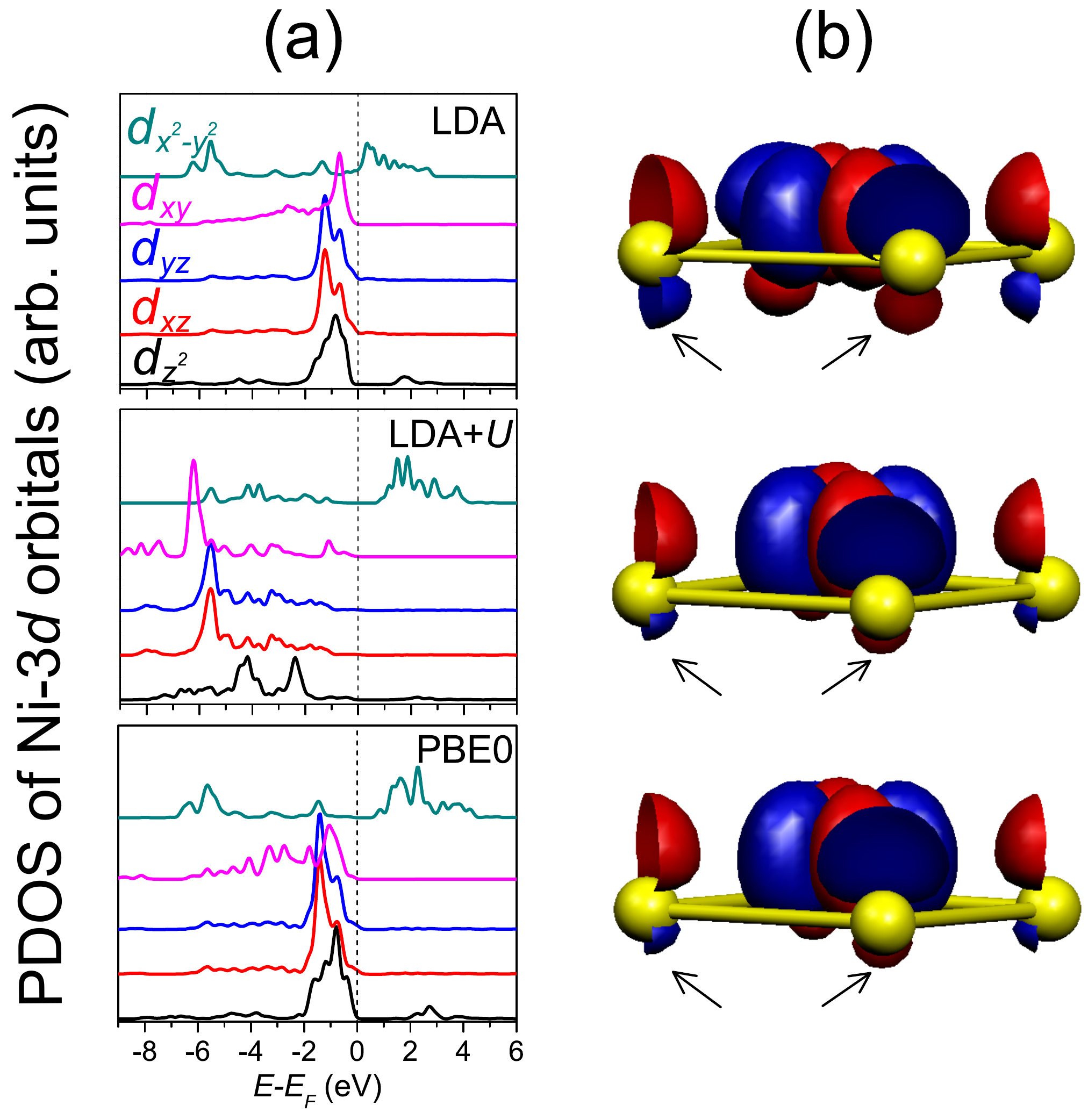}
\caption{(Color online) PDOS for Ni-3$d$ and corresponding wave
function isosurface for Ni-$d_{x^{2}-y^{2}}$ orbital (at $\Gamma$)
in layered Pb$_{2}$NiTiO$_{5}$, calculated using LDA, LDA+$U$ and
PBE0 hybrid density functionals respectively. Compared to LDA,
LDA+$U$ and PBE0 predict smaller downward red (blue)-lobes around
planar O atoms (see arrows), indicating there is more O $s$-$p$
hybridization and a weaker bonding interaction between
Ni-$d_{x^{2}-y^{2}}$ and O-2$p$.} \label{Fig.5}
\end{figure}

Fig.\ 5 (a) shows the PDOS of Ni-3$d$ in layered Ni-PTO, from LDA,
LDA+$U$ and PBE0 calculations. In both the LDA+$U$ and PBE0 PDOS,
the Ni-$d_{x^{2}-y^{2}}$ band lies above $E_{F}$, leading to the
weaker bonding between Ni-$d_{x^{2}-y^{2}}$ and O-2$p$ (Fig.\ 5(b)).
Unlike LDA, which yields a metallic ground state for layered Ni-PTO,
LDA+$U$ and PBE0 predict an intrinsic semiconductor. Since the
LDA+$U$ and PBE0 results agree and are more consistent with the
Ni-$d$ orbital paradigm in diamagnetic insulating SrNiO$_{2}$,
~\cite{Anisimov99p7901} we consider LDA+$U$ and PBE0 to be more
accurate than LDA, and predict that layered Ni-PTO is a
semiconductor.

The calculated PDOS shown in Fig.\ 5 also shows other prominent
features. As a result of the hybridization between Ni and its planar
coordinated O, the occupied Ni-$d$ orbitals, such as $d_{xz}$,
$d_{yz}$ and $d_{z^{2}}$, have a large band width of 2 eV (both LDA
and PBE0 results). As the Ni ion is shifted up, away from the oxygen
plane, the hybridization between Ni-$d_{z^{2}}$ and O-$p_{z}$ can
give rise to bonding and antibonding states. The double peaks in
$d_{xz}$, $d_{yz}$ states should correspond to bonding and
antibonding states, which come from the hybridization between
Ni-$d_{xz}$ ($d_{yz}$) and O-$p_{x}$ ($p_{y}$) (for details see the
supplementary material). The strong hybridization between Ni and
planar O ions indicate the covalency of Ni-O bonds in layered
Ni-PTO.

After exploring the electronic properties of the layered structure,
we extend our investigation to rocksalt ordered Ni-PTO (Fig.\ 6
(a)). In the layered structure, there is a strong {\em bonding}
character between Ni-$d_{x^{2}-y^{2}}$ and O-$p_{z}$ (at $\Gamma$).
However, in rocksalt Ni-PTO, there is an {\em anti-bonding} relation
between Ni-$d_{x^{2}-y^{2}}$ and O-$p_{z}$, while bonding with
O-$p_{z}$ occurs almost exclusively between Ti and O (Fig.\ 6 (b)).
This hybridization pattern indicates that the electron density is
mainly shared between Ti-3$d$ and O-2$p$, and that the
Ni-$d_{x^{2}-y^{2}}$ orbital is almost empty. This leads to the
prediction of a semiconducting ground state for rocksalt ordered
Ni-PTO, even at the LDA level (Fig.\ 6 (c)). A consistent picture
emerges, as LDA+$U$, PBE0 and sc-$GW$ all predict rocksalt
Pb$_{2}$NiTiO$_{5}$ to be a semiconductor.

\begin{figure}[t]
\centering
\includegraphics[width=9cm]{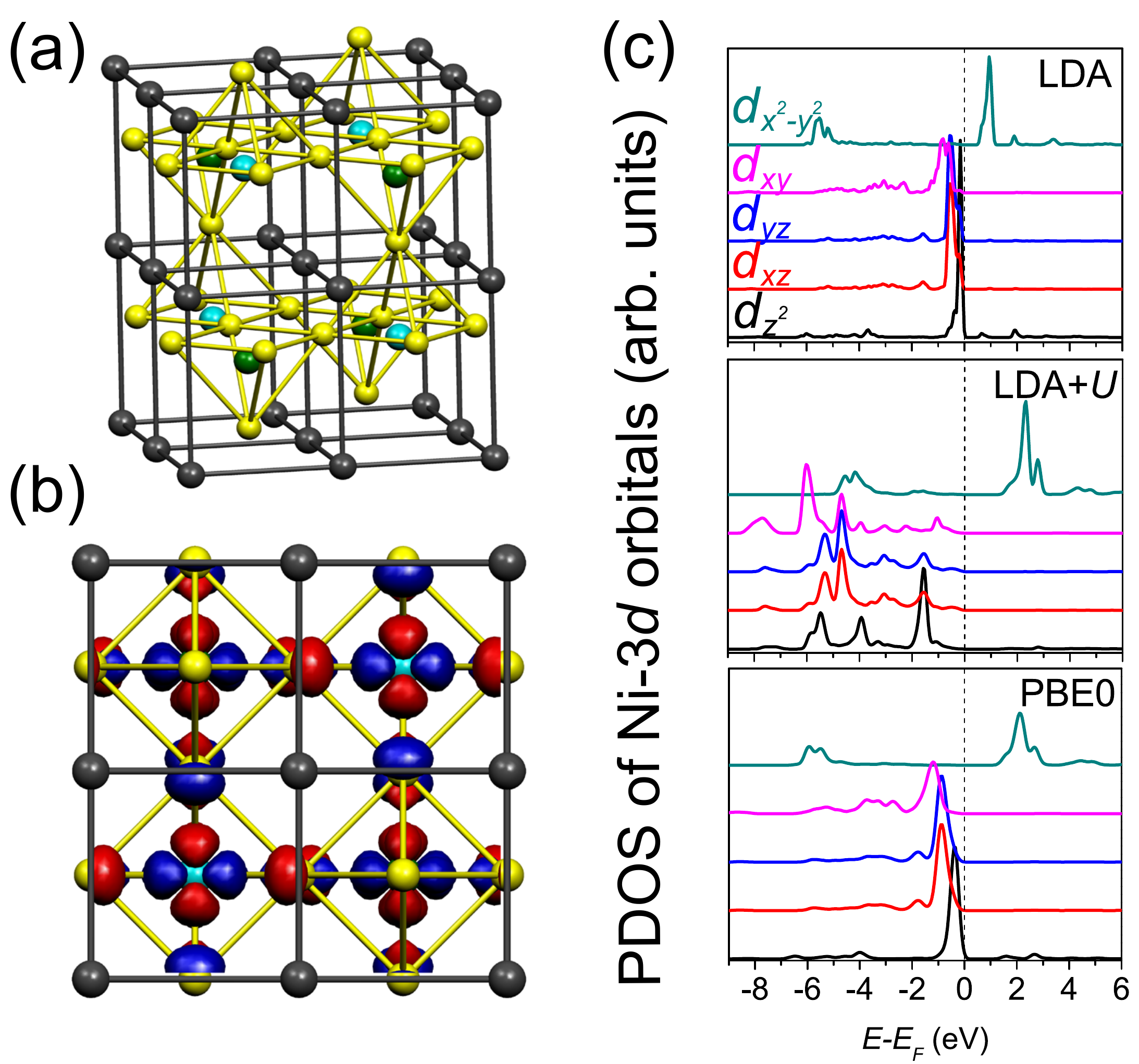}
\caption{(Color online) (a) Atomic structure of rocksalt $B$-site
ordered Pb$_{2}$NiTiO$_{5}$. A 2$\times$2$\times$2 supercell
containing 36 atoms is shown for clarity. (b) Isosurface of LUMO
band (LDA results) at $\Gamma$. It is mainly composed of
$d_{x^{2}-y^{2}}$ orbitals, displaying anti-bonding character
between Ni-3$d_{x^{2}-y^{2}}$ and O-2$p_{z}$. (c) PDOS for Ni-3$d$
orbital, obtained using LDA, LDA+$U$ and PBE0 functional
calculations respectively.} \label{Fig.6}
\end{figure}

Besides the Ni-$d_{x^{2}-y^{2}}$ orbital, other Ni-$d$ bands also
differ between the rocksalt and layered structures of Ni-PTO. In
layered Ni-PTO, a strong hybridization between Ni and O yields
energy bands with a large band-width. However, in rocksalt Ni-PTO,
interrupting the NiO$_{2}$ layer with Ti-O cages makes it more
favorable for the filled O-2$p$ orbitals to mix with the empty 3$d$
orbitals of Ti$^{4+}$~\cite{Wheeler86p2222}. Therefore, Ni-3$d$
orbital is weakly hybridized with O-2$p$, reducing its band width in
rocksalt Ni-PTO (Fig.\ 6 (c)).

\subsection*{\label{sec:level2} $B$-site cation ordering effects}

The results of our electronic band structure analysis indicate that
different $B$-site cation orderings for Pb$_{2}$NiTiO$_{5}$ systems
can give rise to different degrees of Ni-O bond covalency. To
support this, we calculate the Born effective charges (BECs) for
$B$-site cations of Ni-PTO.

Since LDA incorrectly predicts a metallic ground state for layered
Ni-PTO, the BECs and polarization $\vec{P}$ of Ni-PTO should be
calculated at the post-DFT level. The BEC tensor,
$Z^{*}_{\kappa,\alpha\beta}$, is defined as the change in the
macroscopic polarization along direction $\alpha$ in response to a
rigid displacement of a sublattice of atoms $\kappa$ in direction
$\beta$, times the unit cell volume $\Omega_{0}$
~\cite{Ghosez98p6224}:

\begin{equation}
Z^{*}_{\kappa,\alpha\beta}=\Omega_{0}\frac{\partial{P_{\alpha}}}{\partial{\tau_{\kappa,\beta}}}\mid_{\vec{E}=0}
\cong\Omega_{0}\frac{\Delta{P_{\alpha}}}{\Delta{\tau_{\kappa,\beta}}}\mid_{\vec{E}=0}
\end{equation}\

Following this definition, $Z^{*}$ can be estimated by the
finite-difference method: each atom $\kappa$ is displaced slightly
from its equilibrium position by $\Delta{\tau_{\beta}}$, and the
resulting change in polarization $\Delta$$P_{\alpha}$ can be
calculated using the Berry-phase method.
~\cite{Resta94p899,Kingsmith93p1651} The displacement should be
small enough to ensure the validity of the linear treatment of Eq.
(3). For each structure, the $U$ value is determined by a
linear-response calculation, so that LDA+$U$ calculations can
generate an insulating ground state for all Ni-PTO structures. In
this way, we calculate both $Z^{*}$ and $\vec{P}$ at the LDA+$U$
level.

Table III gives the calculated non-zero BECs for Ni ions in Ni-PTO
systems. As Ni$^{2+}$ ion strongly favors a four-fold coordinate
environment, its planar BEC tensor elements ($Z^{*}_{xx}$ and
$Z^{*}_{yy}$) are larger than its nominal charge, indicative of
covalency for the planar Ni-O bonds. Since the apical O moves away
from Ni to create a square planar environment, the bonding between
Ni and the apical O is weak. In the finite-difference simulation of
$Z^{*}$, when Ni ion moves closer to the apical O, the electron
density will transfer from Ni to O. Because of this negative charge
density flow, the Ni ion will have a $Z^{*}_{zz}$ smaller than its
nominal charge. Moreover, the following trend is identified: as
$B$-site cations change from layered to rocksalt ordering, there is
a decrease of the planar BEC tensors of Ni, and an increase of
$Z^{*}$ along Ni-O vacancy direction. This trend is in good
agreement with the bond covalency results we obtained above, that
layered Ni-PTO has a more covalent Ni-O bond and larger $Z^{*}_{xx}$
and $Z^{*}_{yy}$ tensor for Ni, while less covalent Ni-O bond in
rocksalt structure leads to the smaller in-plane Ni BECs.

\begin{table}
\setlength{\tabcolsep}{0.15cm}
\renewcommand{\arraystretch}{1.5}
\caption{The Born effective charges $Z^{*}$ of Ni cations in Ni-PTO
solid solutions. $Z^{*}$ is calculated using LDA+$U$ finite
difference methods.}
\begin{tabular}{c|ccc}
\hline\hline
Composition&Z$^{*}_{xx}$&Z$^{*}_{yy}$&Z$^{*}_{zz}$\\
\hline
1$\times$1$\times$2 Pb$_{2}$NiTiO$_{5}$                              & 2.29& 2.29& 0.44 \\
1$\times$1$\times$3 Pb$_{3}$NiTi$_{2}$O$_{8}$                        & 2.33& 2.33& 0.37 \\
\hline
rocksalt Pb$_{2}$NiTiO$_{5}$                                         & 2.14& 2.14& 0.86 \\
$\sqrt{2}$$\times$$\sqrt{2}$$\times$2 Pb$_{4}$NiTi$_{3}$O$_{11}$     & 2.17& 2.17& 0.72 \\
2$\times$2$\times$2Pb$_{8}$NiTi$_{7}$O$_{23}$                        & 2.15& 2.15& 0.78 \\
\hline\hline
\end{tabular}
\end{table}

In perovskite $AB$O$_{3}$ compounds, a more ionic $B$-O bond usually
yields a larger $E_{\rm g}$, while a more covalent $B$-O bond is
associated with a smaller $E_{\rm g}$. This trend has been verified
for many perovskite systems, such as Pb(Zr$_{x}$Ti$_{1-x}$)O$_{3}$
~\cite{Lee05p094108} and sulfide perovskite BaZrS$_{3}$
~\cite{Bennett09p235115}. In light of this trend, we now examine the
Ni-PTO solid solution systems.

\begin{table*}[bottom]
\setlength{\tabcolsep}{0.075cm}
\renewcommand{\arraystretch}{1.0}
\caption{Reported here are the $c/a$ ratio, polarization $P$ (in
C/m$^{2}$), as well as indirect $E_{\rm g}$ (in eV) calculated using
different methods for all the Ni-PTO solid solutions under
investigation. Also shown are tetragonal PbTiO$_{3}$ results. For
Ni-PTO solid solutions, direct band gaps at $\Gamma$ are given in
parentheses. Structural parameters are obtained by LDA calculations
and, $P$ is calculated using the LDA+$U$ finite difference method.
We use the optimal $\alpha$ for PBE0 calculation.}
\begin{tabular}{c|c|c|c|c|c|c}
\hline\hline Composition    & c/a & $P$ & $E_{\rm g}^{\rm LDA}$ & $E_{\rm g}^{\rm LDA+\emph{U}}$ & $E_{\rm g}^{\rm PBE0}$ & $E_{\rm g}^{\rm sc-\emph{GW}}$  \\
\hline
PbTiO$_{3}$                                    & 1.05   &  0.81 & 1.46(3.05) & 2.17(3.80) & 3.50(4.31) &           \\
\hline
1$\times$1$\times$2 Ni-PTO                     & 1.14   &  0.85 & 0.00(0.00) & 0.70(1.19) & 0.87(1.02) &           \\
1$\times$1$\times$3 Ni-PTO                     & 1.14   &  0.84 & 0.00(0.00) & 0.52(1.57) &            &           \\
\hline
rocksalt Ni-PTO                                & 1.21   &  0.96 & 0.69(1.12) & 1.59(2.65) & 1.71(2.64) & 1.83(2.70)\\
$\sqrt{2}$$\times$$\sqrt{2}$$\times$2 Ni-PTO   & 1.12   &  0.90 & 0.84(0.97) & 1.53(2.24) &            &           \\
2$\times$2$\times$2 Ni-PTO                     & 1.08   &  0.83 & 0.70(0.94) & 1.60(1.90) &            &           \\
\hline\hline
\end{tabular}
\end{table*}

$E_{\rm g}$ of several Ni-PTO solid solutions are summarized in
Table IV. Relative to PTO, $E_{\rm g}$ of Ni-PTO solid solutions are
considerably reduced. For the rocksalt structure, LDA+$U$, PBE0 and
sc-$GW$ calculations increase $E_{\rm g}$ (relative to LDA) by
different amounts, and the trend in calculated $E_{\rm g}$ is:
LDA$<$LDA+$U$$<$PBE0$<$sc-$GW$. Unlike the parameter-free sc-$GW$,
$E_{\rm g}$ results obtained from LDA+$U$ and PBE0 fairly strongly
depend on the computational parameters (exact-exchange mixing
fraction $\alpha$ for PBE0 and Hubbard $U$ term for LDA+$U$). In the
present work, LDA+$U$ and PBE0 $E_{\rm g}$ are in good agreement
with sc-$GW$ results, verifying our parametrization for these
calculations. It is worth noting that even though LDA+$U$ yields
$E_{\rm g}$ similar to PBE0 and sc-$GW$, LDA+$U$ gives a more
localized and perhaps less realistic DOS distribution (Fig.\ 5 (a)
and Fig. 2 in supporting information). Therefore, we only use it to
obtain the relative $E_{\rm g}$ trend, but do not rely on LDA+$U$
for the precise electronic structure. Our calculated $E_{\rm g}$
exhibits obvious $B$-site ordering effects: layered Ni-PTO always
has a smaller $E_{\rm g}$ than rocksalt ordered Ni-PTO, and the
variation in Ni doping percentage has little impact on $E_{\rm g}$
value.

By comparing the BEC tensor and $E_{\rm g}$ results of Ni-PTO solid
solutions, we can establish the relationship between $B$-site cation
ordering and band gap. The electronic properties of Ni-PTO primarily
depends on the Ni-O bonding. In layered Ni-PTO, there is a complete
Ni-O-Ni network, and strong hybridization between Ni-$d$ and O-$p$
increases Ni-O bond covalency, yielding a smaller $E_{\rm g}$.
However, if the Ni-O-Ni network is interrupted by Ti-O bonds, the
hybridization between the filled O-2$p$ orbitals and the empty 3$d$
orbitals of Ti$^{4+}$ will be more favorable~\cite{Wheeler86p2222}.
As a result, there is a more ionic bond between Ni-$d$ and O-$p$,
and the system has a larger $E_{\rm g}$.

Guided by the $E_{\rm g}$ trend obtained above, we estimate that the
lower $E_{\rm g}$ limit of Ni-PTO solid solutions should come from
the layered structure, which could potentially be synthesized via
layer-by-layer deposition techniques; the upper $E_{\rm g}$ limit
corresponds to the more ionic rocksalt ordering structure, which can
perhaps be obtained with a thermal annealing process. According to
our prediction, layered Ni-PTO should have a direct optical band gap
comparable to bulk Si, and $E_{\rm g}$ of rocksalt ordered Ni-PTO is
close to that of BiFeO$_{3}$ ~\cite{Basu08p091905}. In this way,
Ni-PTO solid solutions should have direct band gaps in the visible
light energy range, displaying semiconducting electronic properties.

\section*{\label{sec:level1} Conclusion}

In conclusion, we present theoretical studies on the electronic
properties of Pb(Ni$_{x}$Ti$_{1-x}$)O$_{3-x}$ solid solutions. We
identify the relation between $B$-site cation ordering and $E_{\rm g}$
of Ni-PTO systems. We expect that this relation can be used a guiding
principle to understand $E_{\rm g}$ behaviors for many perovskite
solid solutions. Based on our theoretical calculations, we predict
that the Ni-PTO solid solution can offer both a large polarization and
a direct band gap in the visible light energy range. This combination
of properties indicates that high photovoltaic efficiency can be
realized in solar devices based on Pb(Ni$_{x}$Ti$_{1-x}$)O$_{3-x}$
oxides.

Furthermore, we demonstrate that the standard DFT-LDA provides
qualitatively incorrect electronic structure for layered Ni-PTO, and
potentially other systems being investigated as possible components
in solar harvesting devices. A more accurate treatment of the
Ni-3$d$ states of Ni-PTO required post-DFT LDA calculations to
obtain reasonable band gaps. Of the work presented here, the LDA+$U$
method is the least computationally expensive of the post-DFT
methods, though it sometimes yields unreasonable crystal field
splits and unphysical PDOS features. Though the most computationally
demanding, sc-$GW$ yields band gap results that are closest to
experiment. The band gap results of PBE0 are close to that of
sc-$GW$ and reproduce salient spectral features of the PDOS, while
the computational cost is less than that of $GW$. For the purposes
of calculating accurate band gaps, we suggest that the PBE0 method
is a reasonable compromise of accuracy and cost.

\section*{\label{sec:level1}Acknowledgments}

G. Y. G. and J. W. B. were supported by the Department of Energy
Office of Basic Energy Sciences, under grant number
DE-FG02-07ER46431, and H. T. and A. M. R. by the Office of Naval
Research, under grant number N00014-09-1-0157. The authors would
like to acknowledge K. M. Rabe and D. R. Hamann for helpful
scientific discussions during preparation of the manuscript.
Computational support was provided by a DURIP grant and a Challenge
Grant from the HPCMO.

\bibliography{rappecites}

\begin{thebibliography}{45}
\expandafter\ifx\csname natexlab\endcsname\relax\def\natexlab#1{#1}\fi
\expandafter\ifx\csname bibnamefont\endcsname\relax
  \def\bibnamefont#1{#1}\fi
\expandafter\ifx\csname bibfnamefont\endcsname\relax
  \def\bibfnamefont#1{#1}\fi
\expandafter\ifx\csname citenamefont\endcsname\relax
  \def\citenamefont#1{#1}\fi
\expandafter\ifx\csname url\endcsname\relax
  \def\url#1{\texttt{#1}}\fi
\expandafter\ifx\csname urlprefix\endcsname\relax\def\urlprefix{URL }\fi
\providecommand{\bibinfo}[2]{#2}
\providecommand{\eprint}[2][]{\url{#2}}

\bibitem[{\citenamefont{Yarmarkin et~al.}(2000)\citenamefont{Yarmarkin,
  Gol'tsman, Kazanin, and Lemanov}}]{Yarmarkin00p511}
\bibinfo{author}{\bibfnamefont{V.~K.} \bibnamefont{Yarmarkin}},
  \bibinfo{author}{\bibfnamefont{B.~M.} \bibnamefont{Gol'tsman}},
  \bibinfo{author}{\bibfnamefont{M.~M.} \bibnamefont{Kazanin}},
  \bibnamefont{and} \bibinfo{author}{\bibfnamefont{V.~V.}
  \bibnamefont{Lemanov}}, \bibinfo{journal}{Phys. of Solid State}
  \textbf{\bibinfo{volume}{42}}, \bibinfo{pages}{511} (\bibinfo{year}{2000}).

\bibitem[{\citenamefont{Uprety et~al.}(2007)\citenamefont{Uprety, Ocola, and
  Auciello}}]{Uprety07p084107}
\bibinfo{author}{\bibfnamefont{K.~K.} \bibnamefont{Uprety}},
  \bibinfo{author}{\bibfnamefont{L.~E.} \bibnamefont{Ocola}}, \bibnamefont{and}
  \bibinfo{author}{\bibfnamefont{O.}~\bibnamefont{Auciello}},
  \bibinfo{journal}{J. Appl. Phys.} \textbf{\bibinfo{volume}{102}},
  \bibinfo{pages}{084107} (\bibinfo{year}{2007}).

\bibitem[{\citenamefont{Inoue et~al.}(1986)\citenamefont{Inoue, Sato, Sato, and
  Miyama}}]{Inoue86p2809}
\bibinfo{author}{\bibfnamefont{Y.}~\bibnamefont{Inoue}},
  \bibinfo{author}{\bibfnamefont{K.}~\bibnamefont{Sato}},
  \bibinfo{author}{\bibfnamefont{K.}~\bibnamefont{Sato}}, \bibnamefont{and}
  \bibinfo{author}{\bibfnamefont{H.}~\bibnamefont{Miyama}},
  \bibinfo{journal}{J. Phys. Chem.} \textbf{\bibinfo{volume}{90}},
  \bibinfo{pages}{2809} (\bibinfo{year}{1986}).

\bibitem[{\citenamefont{Glass et~al.}(1974)\citenamefont{Glass, Linde, and
  Negran}}]{Glass74p233}
\bibinfo{author}{\bibfnamefont{A.~M.} \bibnamefont{Glass}},
  \bibinfo{author}{\bibfnamefont{D.~V.~D.} \bibnamefont{Linde}},
  \bibnamefont{and} \bibinfo{author}{\bibfnamefont{T.~J.}
  \bibnamefont{Negran}}, \bibinfo{journal}{Applied Physics Letters}
  \textbf{\bibinfo{volume}{25}}, \bibinfo{pages}{233} (\bibinfo{year}{1974}).

\bibitem[{\citenamefont{Brody}(1973)}]{Brody73p673}
\bibinfo{author}{\bibfnamefont{P.~S.} \bibnamefont{Brody}},
  \bibinfo{journal}{Solid State Communications} \textbf{\bibinfo{volume}{12}},
  \bibinfo{pages}{673} (\bibinfo{year}{1973}).

\bibitem[{\citenamefont{Wang et~al.}(2003)\citenamefont{Wang, Neaton, Zheng,
  Nagarajan, Ogale, Liu, Viehland, Vaithyanathan, Schlom, Waghmare
  et~al.}}]{Wang03p1719}
\bibinfo{author}{\bibfnamefont{J.}~\bibnamefont{Wang}},
  \bibinfo{author}{\bibfnamefont{J.~B.} \bibnamefont{Neaton}},
  \bibinfo{author}{\bibfnamefont{H.}~\bibnamefont{Zheng}},
  \bibinfo{author}{\bibfnamefont{V.}~\bibnamefont{Nagarajan}},
  \bibinfo{author}{\bibfnamefont{S.~B.} \bibnamefont{Ogale}},
  \bibinfo{author}{\bibfnamefont{B.}~\bibnamefont{Liu}},
  \bibinfo{author}{\bibfnamefont{D.}~\bibnamefont{Viehland}},
  \bibinfo{author}{\bibfnamefont{V.}~\bibnamefont{Vaithyanathan}},
  \bibinfo{author}{\bibfnamefont{D.~G.} \bibnamefont{Schlom}},
  \bibinfo{author}{\bibfnamefont{U.~V.} \bibnamefont{Waghmare}},
  \bibnamefont{et~al.}, \bibinfo{journal}{Science}
  \textbf{\bibinfo{volume}{299}}, \bibinfo{pages}{1719} (\bibinfo{year}{2003}).

\bibitem[{\citenamefont{Neaton et~al.}(2005)\citenamefont{Neaton, Ederer,
  Waghmare, Spaldin, and Rabe}}]{Neaton05p014113}
\bibinfo{author}{\bibfnamefont{J.~B.} \bibnamefont{Neaton}},
  \bibinfo{author}{\bibfnamefont{C.}~\bibnamefont{Ederer}},
  \bibinfo{author}{\bibfnamefont{U.~V.} \bibnamefont{Waghmare}},
  \bibinfo{author}{\bibfnamefont{N.~A.} \bibnamefont{Spaldin}},
  \bibnamefont{and} \bibinfo{author}{\bibfnamefont{K.~M.} \bibnamefont{Rabe}},
  \bibinfo{journal}{Phys. Rev. B} \textbf{\bibinfo{volume}{71}},
  \bibinfo{pages}{014113} (\bibinfo{year}{2005}).

\bibitem[{\citenamefont{Basu et~al.}(2008)\citenamefont{Basu, Martin, Chu,
  Gajek, Ramesh, Rai, Xu, and Musfeldt}}]{Basu08p091905}
\bibinfo{author}{\bibfnamefont{S.~R.} \bibnamefont{Basu}},
  \bibinfo{author}{\bibfnamefont{L.~W.} \bibnamefont{Martin}},
  \bibinfo{author}{\bibfnamefont{Y.~H.} \bibnamefont{Chu}},
  \bibinfo{author}{\bibfnamefont{M.}~\bibnamefont{Gajek}},
  \bibinfo{author}{\bibfnamefont{R.}~\bibnamefont{Ramesh}},
  \bibinfo{author}{\bibfnamefont{R.~C.} \bibnamefont{Rai}},
  \bibinfo{author}{\bibfnamefont{X.}~\bibnamefont{Xu}}, \bibnamefont{and}
  \bibinfo{author}{\bibfnamefont{J.~L.} \bibnamefont{Musfeldt}},
  \bibinfo{journal}{Appl. Phys. Lett.} \textbf{\bibinfo{volume}{92}},
  \bibinfo{pages}{091905} (\bibinfo{year}{2008}).

\bibitem[{\citenamefont{Hauser et~al.}(2008)\citenamefont{Hauser, Zhang, Mier,
  Ricciardo, Woodward, Gustafson, Brillson, and Yang}}]{Hauser08p222901}
\bibinfo{author}{\bibfnamefont{A.~J.} \bibnamefont{Hauser}},
  \bibinfo{author}{\bibfnamefont{J.}~\bibnamefont{Zhang}},
  \bibinfo{author}{\bibfnamefont{L.}~\bibnamefont{Mier}},
  \bibinfo{author}{\bibfnamefont{R.~A.} \bibnamefont{Ricciardo}},
  \bibinfo{author}{\bibfnamefont{P.~M.} \bibnamefont{Woodward}},
  \bibinfo{author}{\bibfnamefont{T.~L.} \bibnamefont{Gustafson}},
  \bibinfo{author}{\bibfnamefont{L.~J.} \bibnamefont{Brillson}},
  \bibnamefont{and} \bibinfo{author}{\bibfnamefont{F.~Y.} \bibnamefont{Yang}},
  \bibinfo{journal}{Appl. Phys. Lett.} \textbf{\bibinfo{volume}{92}},
  \bibinfo{pages}{222901} (\bibinfo{year}{2008}).

\bibitem[{\citenamefont{Yang et~al.}(2009)\citenamefont{Yang, Martin, Byrnes,
  Conry, Basu, Paran, Reichertz, Ihlefeld, Adamo, Melville
  et~al.}}]{Yang09p062909}
\bibinfo{author}{\bibfnamefont{S.~Y.} \bibnamefont{Yang}},
  \bibinfo{author}{\bibfnamefont{L.~W.} \bibnamefont{Martin}},
  \bibinfo{author}{\bibfnamefont{S.~J.} \bibnamefont{Byrnes}},
  \bibinfo{author}{\bibfnamefont{T.~E.} \bibnamefont{Conry}},
  \bibinfo{author}{\bibfnamefont{S.~R.} \bibnamefont{Basu}},
  \bibinfo{author}{\bibfnamefont{D.}~\bibnamefont{Paran}},
  \bibinfo{author}{\bibfnamefont{L.}~\bibnamefont{Reichertz}},
  \bibinfo{author}{\bibfnamefont{J.}~\bibnamefont{Ihlefeld}},
  \bibinfo{author}{\bibfnamefont{C.}~\bibnamefont{Adamo}},
  \bibinfo{author}{\bibfnamefont{A.}~\bibnamefont{Melville}},
  \bibnamefont{et~al.}, \bibinfo{journal}{Appl. Phys. Lett.}
  \textbf{\bibinfo{volume}{95}}, \bibinfo{pages}{062909}
  (\bibinfo{year}{2009}).

\bibitem[{\citenamefont{Choi et~al.}(2009)\citenamefont{Choi, Lee, Choi,
  Kiryukhin, and Cheong}}]{Choi09p63}
\bibinfo{author}{\bibfnamefont{T.}~\bibnamefont{Choi}},
  \bibinfo{author}{\bibfnamefont{S.}~\bibnamefont{Lee}},
  \bibinfo{author}{\bibfnamefont{Y.}~\bibnamefont{Choi}},
  \bibinfo{author}{\bibfnamefont{V.}~\bibnamefont{Kiryukhin}},
  \bibnamefont{and} \bibinfo{author}{\bibfnamefont{S.-W.}
  \bibnamefont{Cheong}}, \bibinfo{journal}{Science}
  \textbf{\bibinfo{volume}{324}}, \bibinfo{pages}{63} (\bibinfo{year}{2009}).

\bibitem[{\citenamefont{Bennett et~al.}(2008)\citenamefont{Bennett, Grinberg,
  and Rappe}}]{Bennett08p17409}
\bibinfo{author}{\bibfnamefont{J.~W.} \bibnamefont{Bennett}},
  \bibinfo{author}{\bibfnamefont{I.}~\bibnamefont{Grinberg}}, \bibnamefont{and}
  \bibinfo{author}{\bibfnamefont{A.~M.} \bibnamefont{Rappe}},
  \bibinfo{journal}{J. Am. Chem. Soc.} \textbf{\bibinfo{volume}{130}},
  \bibinfo{pages}{17409} (\bibinfo{year}{2008}).

\bibitem[{\citenamefont{Bennett et~al.}(2010)\citenamefont{Bennett, Grinberg,
  Davies, and Rappe}}]{Bennett10p184106}
\bibinfo{author}{\bibfnamefont{J.~W.} \bibnamefont{Bennett}},
  \bibinfo{author}{\bibfnamefont{I.}~\bibnamefont{Grinberg}},
  \bibinfo{author}{\bibfnamefont{P.~K.} \bibnamefont{Davies}},
  \bibnamefont{and} \bibinfo{author}{\bibfnamefont{A.~M.} \bibnamefont{Rappe}},
  \bibinfo{journal}{Phys. Rev. B} \textbf{\bibinfo{volume}{82}},
  \bibinfo{pages}{184106} (\bibinfo{year}{2010}).

\bibitem[{\citenamefont{Anisimov et~al.}(1991)\citenamefont{Anisimov, Zaanen,
  and Andersen}}]{Anisimov91p943}
\bibinfo{author}{\bibfnamefont{V.~I.} \bibnamefont{Anisimov}},
  \bibinfo{author}{\bibfnamefont{J.}~\bibnamefont{Zaanen}}, \bibnamefont{and}
  \bibinfo{author}{\bibfnamefont{O.~K.} \bibnamefont{Andersen}},
  \bibinfo{journal}{Phys. Rev. B} \textbf{\bibinfo{volume}{44}},
  \bibinfo{pages}{943} (\bibinfo{year}{1991}).

\bibitem[{\citenamefont{Giannozzi et~al.}(2009)\citenamefont{Giannozzi, Baroni,
  Bonini, Calandra, Car, Cavazzoni, Ceresoli, Chiarotti, Cococcioni, Dabo
  et~al.}}]{Giannozzi09p395502}
\bibinfo{author}{\bibfnamefont{P.}~\bibnamefont{Giannozzi}},
  \bibinfo{author}{\bibfnamefont{S.}~\bibnamefont{Baroni}},
  \bibinfo{author}{\bibfnamefont{N.}~\bibnamefont{Bonini}},
  \bibinfo{author}{\bibfnamefont{M.}~\bibnamefont{Calandra}},
  \bibinfo{author}{\bibfnamefont{R.}~\bibnamefont{Car}},
  \bibinfo{author}{\bibfnamefont{C.}~\bibnamefont{Cavazzoni}},
  \bibinfo{author}{\bibfnamefont{D.}~\bibnamefont{Ceresoli}},
  \bibinfo{author}{\bibfnamefont{G.~L.} \bibnamefont{Chiarotti}},
  \bibinfo{author}{\bibfnamefont{M.}~\bibnamefont{Cococcioni}},
  \bibinfo{author}{\bibfnamefont{I.}~\bibnamefont{Dabo}}, \bibnamefont{et~al.},
  \bibinfo{journal}{J. Phys.:Condens. Matter} \textbf{\bibinfo{volume}{21}},
  \bibinfo{pages}{395502} (\bibinfo{year}{2009}).

\bibitem[{\citenamefont{Gonze et~al.}(2002)\citenamefont{Gonze, Beuken,
  Caracas, Detraux, Fuchs, Rignanese, Sindic, Verstraete, Zerah, Jollet
  et~al.}}]{Gonze02p478}
\bibinfo{author}{\bibfnamefont{X.}~\bibnamefont{Gonze}},
  \bibinfo{author}{\bibfnamefont{J.-M.} \bibnamefont{Beuken}},
  \bibinfo{author}{\bibfnamefont{R.}~\bibnamefont{Caracas}},
  \bibinfo{author}{\bibfnamefont{F.}~\bibnamefont{Detraux}},
  \bibinfo{author}{\bibfnamefont{M.}~\bibnamefont{Fuchs}},
  \bibinfo{author}{\bibfnamefont{G.-M.} \bibnamefont{Rignanese}},
  \bibinfo{author}{\bibfnamefont{L.}~\bibnamefont{Sindic}},
  \bibinfo{author}{\bibfnamefont{M.}~\bibnamefont{Verstraete}},
  \bibinfo{author}{\bibfnamefont{G.}~\bibnamefont{Zerah}},
  \bibinfo{author}{\bibfnamefont{F.}~\bibnamefont{Jollet}},
  \bibnamefont{et~al.}, \bibinfo{journal}{Comp. Mater. Sci.}
  \textbf{\bibinfo{volume}{25}}, \bibinfo{pages}{478} (\bibinfo{year}{2002}).

\bibitem[{\citenamefont{Monkhorst and Pack}(1976)}]{Monkhorst76p5188}
\bibinfo{author}{\bibfnamefont{H.~J.} \bibnamefont{Monkhorst}}
  \bibnamefont{and} \bibinfo{author}{\bibfnamefont{J.~D.} \bibnamefont{Pack}},
  \bibinfo{journal}{Phys. Rev. B} \textbf{\bibinfo{volume}{13}},
  \bibinfo{pages}{5188} (\bibinfo{year}{1976}).

\bibitem[{\citenamefont{Rappe et~al.}(1990)\citenamefont{Rappe, Rabe, Kaxiras,
  and Joannopoulos}}]{Rappe90p1227}
\bibinfo{author}{\bibfnamefont{A.~M.} \bibnamefont{Rappe}},
  \bibinfo{author}{\bibfnamefont{K.~M.} \bibnamefont{Rabe}},
  \bibinfo{author}{\bibfnamefont{E.}~\bibnamefont{Kaxiras}}, \bibnamefont{and}
  \bibinfo{author}{\bibfnamefont{J.~D.} \bibnamefont{Joannopoulos}},
  \bibinfo{journal}{Phys. Rev. B Rapid Comm.} \textbf{\bibinfo{volume}{41}},
  \bibinfo{pages}{1227} (\bibinfo{year}{1990}).

\bibitem[{\citenamefont{Ramer and Rappe}(1999)}]{Ramer99p12471}
\bibinfo{author}{\bibfnamefont{N.~J.} \bibnamefont{Ramer}} \bibnamefont{and}
  \bibinfo{author}{\bibfnamefont{A.~M.} \bibnamefont{Rappe}},
  \bibinfo{journal}{Phys. Rev. B} \textbf{\bibinfo{volume}{59}},
  \bibinfo{pages}{12471} (\bibinfo{year}{1999}).

\bibitem[{Opi()}]{Opium}
\bibinfo{howpublished}{http://opium.sourceforge.net}.

\bibitem[{\citenamefont{Resta}(1994)}]{Resta94p899}
\bibinfo{author}{\bibfnamefont{R.}~\bibnamefont{Resta}}, \bibinfo{journal}{Rev.
  Mod. Phys.} \textbf{\bibinfo{volume}{66}}, \bibinfo{pages}{899}
  (\bibinfo{year}{1994}).

\bibitem[{\citenamefont{King-Smith and Vanderbilt}(1993)}]{Kingsmith93p1651}
\bibinfo{author}{\bibfnamefont{R.~D.} \bibnamefont{King-Smith}}
  \bibnamefont{and}
  \bibinfo{author}{\bibfnamefont{D.}~\bibnamefont{Vanderbilt}},
  \bibinfo{journal}{Phys. Rev. B} \textbf{\bibinfo{volume}{47}},
  \bibinfo{pages}{1651} (\bibinfo{year}{1993}).

\bibitem[{\citenamefont{Hybertsen and Louie}(1985)}]{Hybertsen85p1418}
\bibinfo{author}{\bibfnamefont{M.~S.} \bibnamefont{Hybertsen}}
  \bibnamefont{and} \bibinfo{author}{\bibfnamefont{S.~G.} \bibnamefont{Louie}},
  \bibinfo{journal}{Phys. Rev. Lett.} \textbf{\bibinfo{volume}{55}},
  \bibinfo{pages}{1418} (\bibinfo{year}{1985}).

\bibitem[{\citenamefont{Hybertsen and Louie}(1986)}]{Hybertsen86p5390}
\bibinfo{author}{\bibfnamefont{M.~S.} \bibnamefont{Hybertsen}}
  \bibnamefont{and} \bibinfo{author}{\bibfnamefont{S.~G.} \bibnamefont{Louie}},
  \bibinfo{journal}{Phys. Rev. B} \textbf{\bibinfo{volume}{34}},
  \bibinfo{pages}{5390} (\bibinfo{year}{1986}).

\bibitem[{\citenamefont{Cococcioni and
  de~Gironcoli}(2005)}]{Cococcioni05p035105}
\bibinfo{author}{\bibfnamefont{M.}~\bibnamefont{Cococcioni}} \bibnamefont{and}
  \bibinfo{author}{\bibfnamefont{S.}~\bibnamefont{de~Gironcoli}},
  \bibinfo{journal}{Phys. Rev. B} \textbf{\bibinfo{volume}{71}},
  \bibinfo{pages}{035105} (\bibinfo{year}{2005}).

\bibitem[{\citenamefont{Perdew et~al.}(1996{\natexlab{a}})\citenamefont{Perdew,
  Ernzerhof, and Burke}}]{Perdew96p9982}
\bibinfo{author}{\bibfnamefont{J.~P.} \bibnamefont{Perdew}},
  \bibinfo{author}{\bibfnamefont{M.}~\bibnamefont{Ernzerhof}},
  \bibnamefont{and} \bibinfo{author}{\bibfnamefont{K.}~\bibnamefont{Burke}},
  \bibinfo{journal}{J. Chem. Phys.} \textbf{\bibinfo{volume}{105}},
  \bibinfo{pages}{9982} (\bibinfo{year}{1996}{\natexlab{a}}).

\bibitem[{\citenamefont{Perdew et~al.}(1996{\natexlab{b}})\citenamefont{Perdew,
  Burke, and Ernzerhof}}]{Perdew96p3865}
\bibinfo{author}{\bibfnamefont{J.~P.} \bibnamefont{Perdew}},
  \bibinfo{author}{\bibfnamefont{K.}~\bibnamefont{Burke}}, \bibnamefont{and}
  \bibinfo{author}{\bibfnamefont{M.}~\bibnamefont{Ernzerhof}},
  \bibinfo{journal}{Phys. Rev. Lett.} \textbf{\bibinfo{volume}{77}},
  \bibinfo{pages}{3865} (\bibinfo{year}{1996}{\natexlab{b}}).

\bibitem[{\citenamefont{Paier et~al.}(2006)\citenamefont{Paier, Marsman,
  Hummer, Kresse, Gerber, and Angyan}}]{Paier06p154709}
\bibinfo{author}{\bibfnamefont{J.}~\bibnamefont{Paier}},
  \bibinfo{author}{\bibfnamefont{M.}~\bibnamefont{Marsman}},
  \bibinfo{author}{\bibfnamefont{K.}~\bibnamefont{Hummer}},
  \bibinfo{author}{\bibfnamefont{G.}~\bibnamefont{Kresse}},
  \bibinfo{author}{\bibfnamefont{I.~C.} \bibnamefont{Gerber}},
  \bibnamefont{and} \bibinfo{author}{\bibfnamefont{J.~G.}
  \bibnamefont{Angyan}}, \bibinfo{journal}{J. Chem. Phys.}
  \textbf{\bibinfo{volume}{124}}, \bibinfo{pages}{154709}
  (\bibinfo{year}{2006}).

\bibitem[{\citenamefont{Alkauskas et~al.}(2008)\citenamefont{Alkauskas,
  Broqvist, Devynck, and Pasquarello}}]{Alkauskas08p106802}
\bibinfo{author}{\bibfnamefont{A.}~\bibnamefont{Alkauskas}},
  \bibinfo{author}{\bibfnamefont{P.}~\bibnamefont{Broqvist}},
  \bibinfo{author}{\bibfnamefont{F.}~\bibnamefont{Devynck}}, \bibnamefont{and}
  \bibinfo{author}{\bibfnamefont{A.}~\bibnamefont{Pasquarello}},
  \bibinfo{journal}{Phys. Rev. Lett.} \textbf{\bibinfo{volume}{101}},
  \bibinfo{pages}{106802} (\bibinfo{year}{2008}).

\bibitem[{\citenamefont{Shirane et~al.}(1956)\citenamefont{Shirane, Pepinsky,
  and Frazer}}]{Shirane56p131}
\bibinfo{author}{\bibfnamefont{G.}~\bibnamefont{Shirane}},
  \bibinfo{author}{\bibfnamefont{R.}~\bibnamefont{Pepinsky}}, \bibnamefont{and}
  \bibinfo{author}{\bibfnamefont{B.~C.} \bibnamefont{Frazer}},
  \bibinfo{journal}{Acta Cryst.} \textbf{\bibinfo{volume}{9}},
  \bibinfo{pages}{131} (\bibinfo{year}{1956}).

\bibitem[{\citenamefont{Cheetham and Hope}(1983)}]{Cheetham83p6964}
\bibinfo{author}{\bibfnamefont{A.~K.} \bibnamefont{Cheetham}} \bibnamefont{and}
  \bibinfo{author}{\bibfnamefont{D.~A.~O.} \bibnamefont{Hope}},
  \bibinfo{journal}{Phys. Rev. B} \textbf{\bibinfo{volume}{27}},
  \bibinfo{pages}{6964} (\bibinfo{year}{1983}).

\bibitem[{\citenamefont{Sawatzky and Allen}(1984)}]{Sawatzky84p2339}
\bibinfo{author}{\bibfnamefont{G.~A.} \bibnamefont{Sawatzky}} \bibnamefont{and}
  \bibinfo{author}{\bibfnamefont{J.~W.} \bibnamefont{Allen}},
  \bibinfo{journal}{Phys. Rev. Lett.} \textbf{\bibinfo{volume}{53}},
  \bibinfo{pages}{2339} (\bibinfo{year}{1984}).

\bibitem[{\citenamefont{Hedin}(1965)}]{Hedin65p796}
\bibinfo{author}{\bibfnamefont{L.}~\bibnamefont{Hedin}},
  \bibinfo{journal}{Phys. Rev.} \textbf{\bibinfo{volume}{139}},
  \bibinfo{pages}{A796} (\bibinfo{year}{1965}).

\bibitem[{\citenamefont{Lebegue et~al.}(2003)\citenamefont{Lebegue, Arnaud,
  Alouani, and Bloechl}}]{Lebegue03p155208}
\bibinfo{author}{\bibfnamefont{S.}~\bibnamefont{Lebegue}},
  \bibinfo{author}{\bibfnamefont{B.}~\bibnamefont{Arnaud}},
  \bibinfo{author}{\bibfnamefont{M.}~\bibnamefont{Alouani}}, \bibnamefont{and}
  \bibinfo{author}{\bibfnamefont{P.~E.} \bibnamefont{Bloechl}},
  \bibinfo{journal}{Phys. Rev. B} \textbf{\bibinfo{volume}{67}},
  \bibinfo{pages}{155208} (\bibinfo{year}{2003}).

\bibitem[{\citenamefont{Bruneval et~al.}(2006)\citenamefont{Bruneval, Vast, and
  Reining}}]{Bruneval06p045102}
\bibinfo{author}{\bibfnamefont{F.}~\bibnamefont{Bruneval}},
  \bibinfo{author}{\bibfnamefont{N.}~\bibnamefont{Vast}}, \bibnamefont{and}
  \bibinfo{author}{\bibfnamefont{L.}~\bibnamefont{Reining}},
  \bibinfo{journal}{Phys. Rev. B} \textbf{\bibinfo{volume}{74}},
  \bibinfo{pages}{045102} (\bibinfo{year}{2006}).

\bibitem[{\citenamefont{Anisimov}(2000)}]{Anisimovbook}
\bibinfo{author}{\bibfnamefont{V.~I.} \bibnamefont{Anisimov}},
  \emph{\bibinfo{title}{Strong coulomb correlations in electronic structure
  calculations}} (\bibinfo{publisher}{Gordon and Breach Science Publishers},
  \bibinfo{year}{2000}).

\bibitem[{\citenamefont{Schuler et~al.}(2005)\citenamefont{Schuler, Ederer,
  Itza-Ortiz, Woods, Callcott, and Woicik}}]{Schuler05p115113}
\bibinfo{author}{\bibfnamefont{T.~M.} \bibnamefont{Schuler}},
  \bibinfo{author}{\bibfnamefont{D.~L.} \bibnamefont{Ederer}},
  \bibinfo{author}{\bibfnamefont{S.}~\bibnamefont{Itza-Ortiz}},
  \bibinfo{author}{\bibfnamefont{G.~T.} \bibnamefont{Woods}},
  \bibinfo{author}{\bibfnamefont{T.~A.} \bibnamefont{Callcott}},
  \bibnamefont{and} \bibinfo{author}{\bibfnamefont{J.~C.}
  \bibnamefont{Woicik}}, \bibinfo{journal}{Phys. Rev. B}
  \textbf{\bibinfo{volume}{71}}, \bibinfo{pages}{115113}
  (\bibinfo{year}{2005}).

\bibitem[{\citenamefont{Kunes et~al.}(2007)\citenamefont{Kunes, Anisimov,
  Skornyakov, Lukoyanov, and Vollhardt}}]{Kunes07p156404}
\bibinfo{author}{\bibfnamefont{J.}~\bibnamefont{Kunes}},
  \bibinfo{author}{\bibfnamefont{V.~I.} \bibnamefont{Anisimov}},
  \bibinfo{author}{\bibfnamefont{S.~L.} \bibnamefont{Skornyakov}},
  \bibinfo{author}{\bibfnamefont{A.~V.} \bibnamefont{Lukoyanov}},
  \bibnamefont{and}
  \bibinfo{author}{\bibfnamefont{D.}~\bibnamefont{Vollhardt}},
  \bibinfo{journal}{Phys. Rev. Lett.} \textbf{\bibinfo{volume}{99}},
  \bibinfo{pages}{156404} (\bibinfo{year}{2007}).

\bibitem[{\citenamefont{\"{o}ger and Vink}(1956)}]{Kroger56p273}
\bibinfo{author}{\bibfnamefont{F.~A.~K.} \bibnamefont{\"{o}ger}}
  \bibnamefont{and} \bibinfo{author}{\bibfnamefont{H.~J.} \bibnamefont{Vink}},
  \bibinfo{journal}{Solid State Physics} \textbf{\bibinfo{volume}{101}},
  \bibinfo{pages}{273} (\bibinfo{year}{1956}).

\bibitem[{\citenamefont{Anisimov et~al.}(1999)\citenamefont{Anisimov,
  Bukhvalov, and Rice}}]{Anisimov99p7901}
\bibinfo{author}{\bibfnamefont{V.~I.} \bibnamefont{Anisimov}},
  \bibinfo{author}{\bibfnamefont{D.}~\bibnamefont{Bukhvalov}},
  \bibnamefont{and} \bibinfo{author}{\bibfnamefont{T.~M.} \bibnamefont{Rice}},
  \bibinfo{journal}{Phys. Rev. B.} \textbf{\bibinfo{volume}{59}},
  \bibinfo{pages}{7901} (\bibinfo{year}{1999}).

\bibitem[{\citenamefont{Bilc et~al.}(2008)\citenamefont{Bilc, Orlando, Shaltaf,
  Rignanese, Iniguez, and Ghosez}}]{Blic08p165107}
\bibinfo{author}{\bibfnamefont{D.~I.} \bibnamefont{Bilc}},
  \bibinfo{author}{\bibfnamefont{R.}~\bibnamefont{Orlando}},
  \bibinfo{author}{\bibfnamefont{R.}~\bibnamefont{Shaltaf}},
  \bibinfo{author}{\bibfnamefont{G.-M.} \bibnamefont{Rignanese}},
  \bibinfo{author}{\bibfnamefont{J.}~\bibnamefont{Iniguez}}, \bibnamefont{and}
  \bibinfo{author}{\bibfnamefont{P.}~\bibnamefont{Ghosez}},
  \bibinfo{journal}{Phys. Rev. B} \textbf{\bibinfo{volume}{77}},
  \bibinfo{pages}{165107} (\bibinfo{year}{2008}).

\bibitem[{\citenamefont{Wheeler et~al.}(1986)\citenamefont{Wheeler, Whangbo,
  Hughbanks, Hoffmann, Burdett, and Albrightl}}]{Wheeler86p2222}
\bibinfo{author}{\bibfnamefont{R.~A.} \bibnamefont{Wheeler}},
  \bibinfo{author}{\bibfnamefont{M.~H.} \bibnamefont{Whangbo}},
  \bibinfo{author}{\bibfnamefont{T.}~\bibnamefont{Hughbanks}},
  \bibinfo{author}{\bibfnamefont{R.}~\bibnamefont{Hoffmann}},
  \bibinfo{author}{\bibfnamefont{J.~K.} \bibnamefont{Burdett}},
  \bibnamefont{and} \bibinfo{author}{\bibfnamefont{T.~A.}
  \bibnamefont{Albrightl}}, \bibinfo{journal}{J. Am. Chem. Soc.}
  \textbf{\bibinfo{volume}{108}}, \bibinfo{pages}{2222} (\bibinfo{year}{1986}).

\bibitem[{\citenamefont{Ghosez et~al.}(1998)\citenamefont{Ghosez, Michenaud,
  and Gonze}}]{Ghosez98p6224}
\bibinfo{author}{\bibfnamefont{P.}~\bibnamefont{Ghosez}},
  \bibinfo{author}{\bibfnamefont{J.-P.} \bibnamefont{Michenaud}},
  \bibnamefont{and} \bibinfo{author}{\bibfnamefont{X.}~\bibnamefont{Gonze}},
  \bibinfo{journal}{Phys. Rev. B} \textbf{\bibinfo{volume}{58}},
  \bibinfo{pages}{6224} (\bibinfo{year}{1998}).

\bibitem[{\citenamefont{Lee et~al.}(2005)\citenamefont{Lee, Kang, Cho, Xiao,
  Morko\c{c}, Kang, Lee, Li, Wei, Snyder et~al.}}]{Lee05p094108}
\bibinfo{author}{\bibfnamefont{H.}~\bibnamefont{Lee}},
  \bibinfo{author}{\bibfnamefont{Y.~S.} \bibnamefont{Kang}},
  \bibinfo{author}{\bibfnamefont{S.-J.} \bibnamefont{Cho}},
  \bibinfo{author}{\bibfnamefont{B.}~\bibnamefont{Xiao}},
  \bibinfo{author}{\bibfnamefont{H.}~\bibnamefont{Morko\c{c}}},
  \bibinfo{author}{\bibfnamefont{T.~D.} \bibnamefont{Kang}},
  \bibinfo{author}{\bibfnamefont{G.~S.} \bibnamefont{Lee}},
  \bibinfo{author}{\bibfnamefont{J.}~\bibnamefont{Li}},
  \bibinfo{author}{\bibfnamefont{S.-H.} \bibnamefont{Wei}},
  \bibinfo{author}{\bibfnamefont{P.~G.} \bibnamefont{Snyder}},
  \bibnamefont{et~al.}, \bibinfo{journal}{J. Appl. Phys.}
  \textbf{\bibinfo{volume}{98}}, \bibinfo{pages}{094108}
  (\bibinfo{year}{2005}).

\bibitem[{\citenamefont{Bennett et~al.}(2009)\citenamefont{Bennett, Grinberg,
  and Rappe}}]{Bennett09p235115}
\bibinfo{author}{\bibfnamefont{J.~W.} \bibnamefont{Bennett}},
  \bibinfo{author}{\bibfnamefont{I.}~\bibnamefont{Grinberg}}, \bibnamefont{and}
  \bibinfo{author}{\bibfnamefont{A.~M.} \bibnamefont{Rappe}},
  \bibinfo{journal}{Phys. Rev. B} \textbf{\bibinfo{volume}{79}},
  \bibinfo{pages}{235115} (\bibinfo{year}{2009}).

\end{thebibliography}

\end{document}



\title{\bf{Supplementary materials - Post density functional theoretical studies of highly
polar semiconducting Pb(Ti$_{1-x}$Ni$_{x}$)O$_{3-x}$ solid
solutions: The effects of cation arrangement on band gap} \\[11pt] }

\author{G. Y. Gou$^{1}$, J. W. Bennett$^{2}$, H. Takenaka$^{1}$ and A. M. Rappe$^{1}$}
\affiliation{
  1. The Makineni Theoretical Laboratories, Department of
  Chemistry, University of Pennsylvania, Philadelphia, PA
  19104-6323, USA\\
  2. Department of Physics and Astronomy, Rutgers University, Piscataway, New Jersey 08854-8019, USA\\}

\maketitle

\begin{figure*}[t]
\centering
\includegraphics[width=11.0cm]{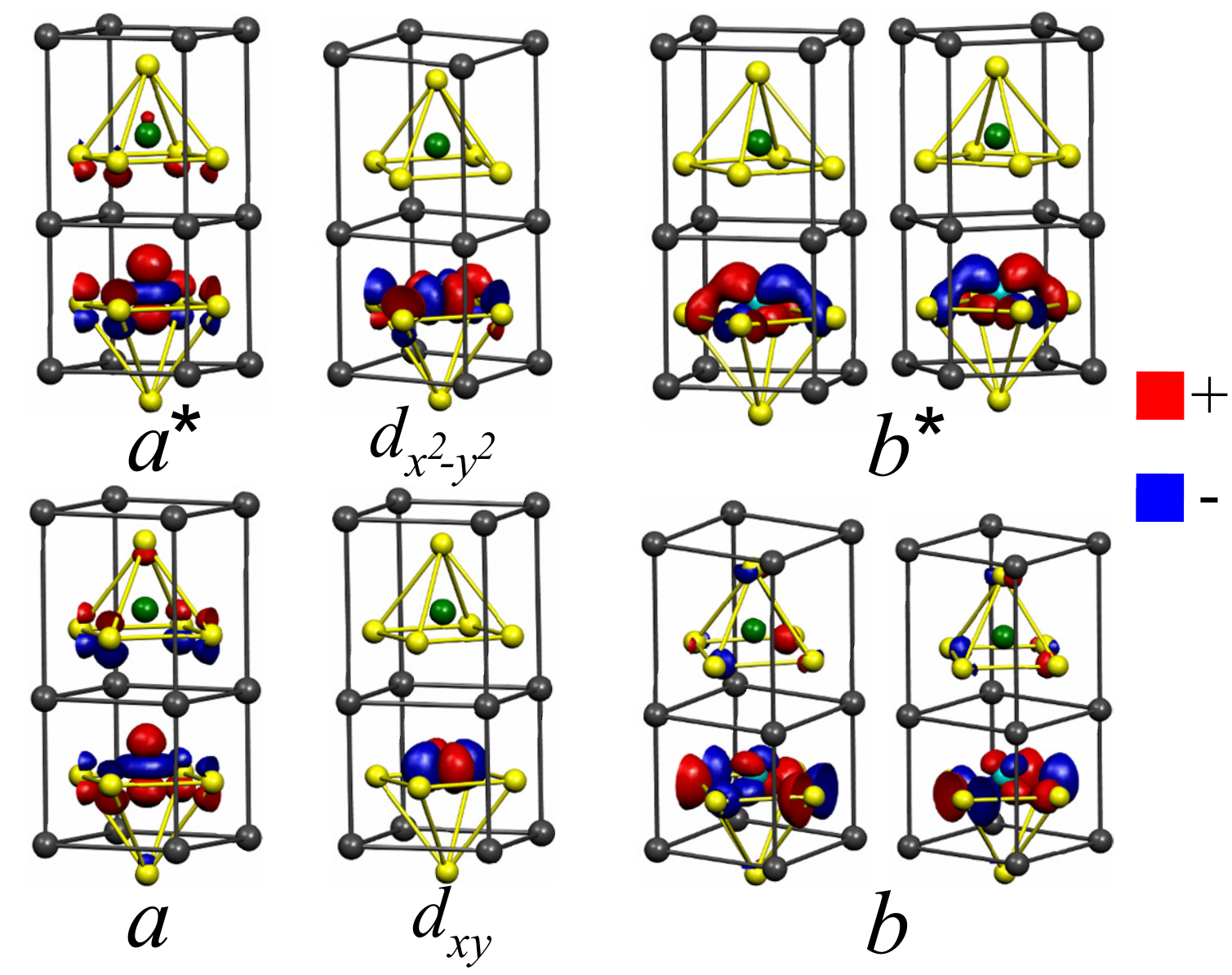}
\caption{(Color online) The wave function isosurfaces of the Ni-3$d$
orbitals at $\Gamma$ from the layered Pb$_{2}$NiTiO$_{5}$,
calculated by LDA. The isosurface value is $\pm$ 0.04 e/\AA$^{3}$.
Except for the non-bonding Ni-3$d_{xy}$ state, the wave functions
show obvious hybridization between Ni-3$d$ and O-2$p$. $a$ and
$a^{*}$ correspond to the $\pi$-type bonding and antibonding states
between Ni-3$d_{z^{2}}$ and O-$p_{z}$. $b$ and $b^{*}$ are
degenerate bonding and antibonding states from Ni-$d_{xz}$
($d_{yz}$) and O-$p_{x}$($p_{y}$).}\label{Fig.1}
\end{figure*}

\begin{figure*}[t]
\centering
\includegraphics[width=9.0cm]{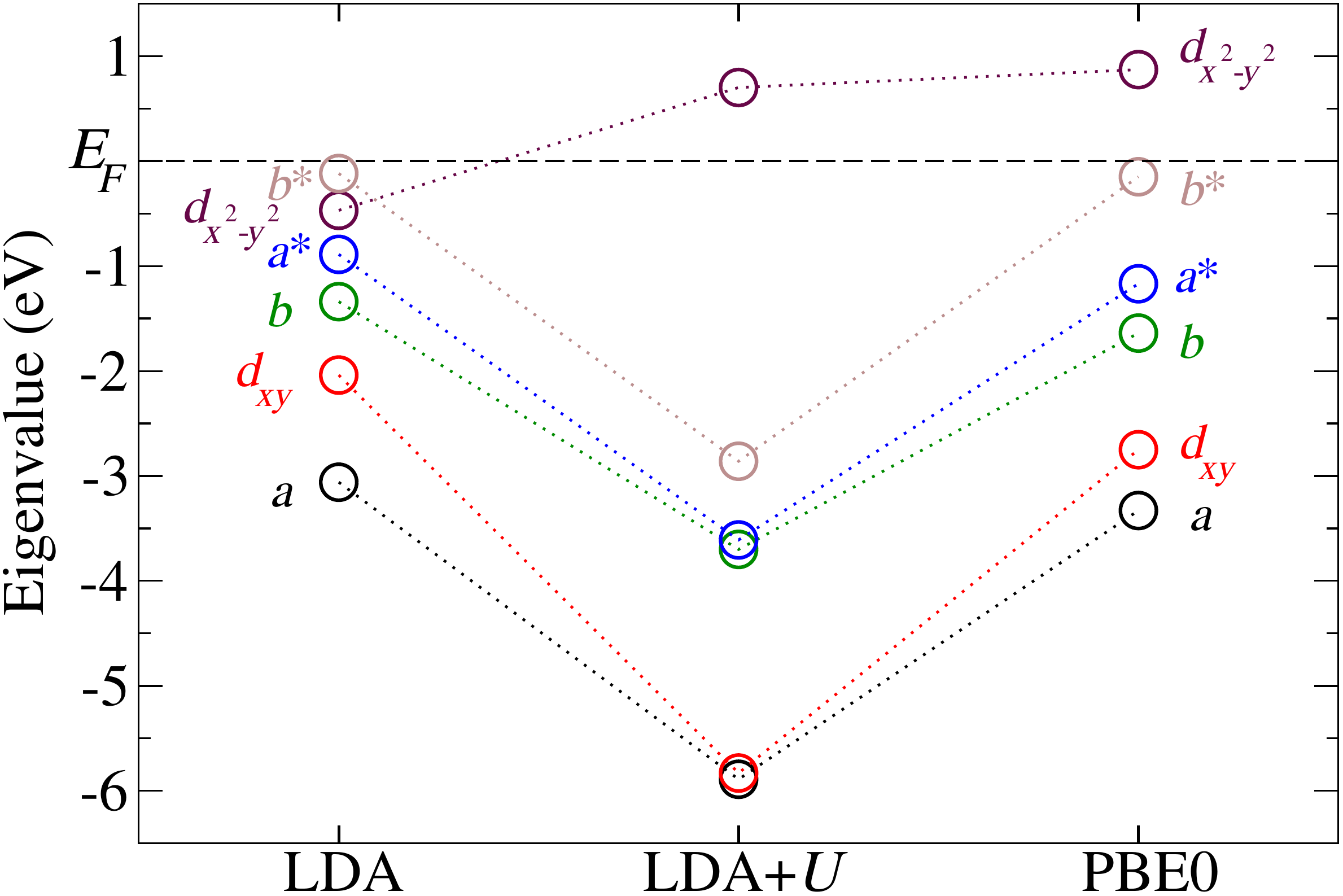}
\caption{(Color online) The energy levels (in eV) for the
corresponding eigen states in Fig. 1, calculated by LDA, PBE0 and
LDA+$U$ methods. Fermi energy level is taken as energy zero. Both
LDA+$U$ and PBE0 predict an empty Ni-$d_{x^{2}-y^{2}}$ state.
However, LDA+$U$ predicts energy levels of occupied Ni-$d$ states
far below the Fermi level, leading to the overall improper PDOS
feature (Fig. 5 (a) in the main text) for layered Ni-PTO. Note that
PBE0 has similar PDOS to LDA, and has $d_{x^{2}-y^{2}}$ unfilled
above $E_{F}$.}\label{Fig.2}
\end{figure*}